# A new paradigm for accelerating clinical data science at Stanford Medicine

Somalee Datta[1], Jose Posada[1,2], Garrick Olson[1], Wencheng Li[1], Ciaran O'Reilly[1], Deepa Balraj[1], Joe Mesterhazy[1], Joe Pallas[1], Priyamvada Desai[1], Nigam H. Shah[1,2]

1. Research IT, Technology & Digital Solutions, Stanford Medicine.
2. Stanford Center for Biomedical Informatics Research, School of Medicine, Stanford University

## Abstract:

Stanford Medicine is building a new data platform for our academic research community to do better clinical data science. Hospitals have a large amount of patient data and researchers have demonstrated the ability to reuse that data and AI approaches to derive novel insights, support patient care, and improve care quality. However, the traditional data warehouse and Honest Broker approaches that are in current use, are not scalable. We are establishing a new secure Big Data platform that aims to reduce time to access and analyze data. In this platform, data is anonymized to preserve patient data privacy and made available preparatory to Institutional Review Board (IRB) submission. Furthermore, the data is standardized such that analysis done at Stanford can be replicated elsewhere using the same analytical code and clinical concepts. Finally, the analytics data warehouse integrates with a secure data science computational facility to support large scale data analytics. The ecosystem is designed to bring the modern data science community to highly sensitive clinical data in a secure and collaborative big data analytics environment with a goal to enable bigger, better and faster science.

## Background:

Healthcare is at a nexus where the data generated during the course of clinical care or research can be used to advance both the science and practice of medicine. With the widespread adoption of electronic health records (EHR), driven by the Health Information Technology for Economic and Clinical Health (HITECH) Act of 2009 [Gold2016], most academic centers now have more than ten years of EHR data available for reuse. [Lowe2009, Evans2012, Danciu2014, Kunjan2015, Turley2016, Foran2017]. In addition, hospitals have a fast growing amount of data from sources such as radiology, pathology, and bedside monitoring that are ripe for reuse to improve patient care and to advance scientific discovery [Shah2019, Rajkomar2019]. A sampling of research in recent years at Stanford [Banda2019, Banerjee2019, Dunnmon2019, Gombar2019, Hernandez-Boussard2016, Hernandez-Boussard2019, Jung2019, Parthipan2019, Rajpurkar2018, Ross2019, Tamang2015, Tamang2017a, Tamang2017b, Wang2018] suggests that researchers are able to reuse the collected data to derive novel insights, support patient care, and improve care quality using Artificial Intelligence (AI) approaches. Such efforts can improve the standard of care today [Gombar2019, Jung2019] and also hold the promise to improve health outcomes and reduce cost of care in the coming years [Tamang2017b].

Stanford's efforts to explore the use of AI in Medicine (AIM) started in the 1970s with the Stanford University Medical EXperimental computer for Artificial Intelligence in Medicine (SUMEX-AIM) project [Kulikowski2019], a national computer resource (1973-1992) funded by NIH to promote applications of AI to biological and medical problems. Over the past decade, Stanford School of Medicine Dean's Office has made significant investment in support of a clinical data ecosystem for use for research. Stanford's first generation Clinical Data Warehouse (CDW), STRIDE - Stanford Translational Research Integrated Database Environment [Lowe2009], has been serving Stanford research community since 2008.

STRIDE has an associated cohort building tool, Anonymous Patient Cohort Discovery Tool [Lowe2009], that returns patient counts matching a given query and is used to design studies prior to Institutional Review Board (IRB) submission. The tool is designed to answer the question "How many patients in the CDW contain these attributes?" No individual patient data is exposed during cohort search. A chart review tool, accessible after IRB approval [Grady2015], allows rapid view of the retrieved patient cohort. In addition, there is a real time Complex Event Processing (CEP) engine [Weber2010], that uses HL7 feeds to identify potential research participants.

Stanford's Honest Broker [Choi2015], in part funded by Dean's Office and in part funded by the National Center for Advancing Translational Sciences (NCATS) Clinical and Translational Science Awards (CTSA), is composed of two teams. The Research IT team (RIT) at Technology & Digital Solutions has performed the role of data stewards since the release of STRIDE in 2008. RIT transfers data from our hospitals, normalizes the data for research use, develops Cohort and Chart review tools, develops de-identification pipelines and supports end users. The Research Informatics Center (RIC) in the Dean's Office represents the analytics team and was split up from RIT in 2017 and supports custom query development and extraction. The Honest Broker teams have provided over 5000 clinical data consults since 2008 resulting in hundreds of publications, including many that employ machine learning and other forms of AI techniques. The numbers have increased year over year. In 2019, approximately 650 studies involving 530 PIs and 1500 researchers were approved by IRB. Approximately 58% of the approved IRBs could use the existing STRIDE cohort and chart review tools in a self-service manner. The Honest Broker teams served over 800 consultations including both pre and post-IRB support.

## Vision:

There is an appetite for diverse data types at Stanford as illustrated by a sampling of publications, *e.g.*, radiology [Dunnmon2019, Patel2020], echocardiography [Dykes2019], and bedside monitoring [Miller2018]. Our Hospitals are currently accruing research usable data at a petabyte scale. For example, the current rate of growth for radiology and echo data is estimated at 450 TB/yr, bedside monitoring is estimated at 100 TB/yr, digital pathology at 1000 terabytes/yr. Research IT is starting to bring these datasets in a new STAnford medicine Research data Repository (STARR, http://starr.stanford.edu) with the intent of a) making these rich but often proprietary sources of data "analysis-ready", b) linking the data to Clarity for "completeness", and c) bringing in ancillary data sources (e.g. PowerPath for digital pathology) to make the data metadata "rich". Our intent is to support the necessary data warehouses, data models, identified or anonymized non-human subject data, and analytical tools based on the data in STARR.

We also envision a world where we can bring more scientists to data, do better team science while preserving patient privacy, fail fast and innovate faster, and bring innovations to patients in a timely fashion.

## Motivation:

Our existing ecosystem of STRIDE CDW, Cohort and Chart review tools, and our Honest Broker services are unable to keep up with growing demand for greater volume and complexity of data. Researchers are waiting longer for IRB approval, privacy, and security reviews, and Honest Broker services.  In this section, we motivate our new clinical data science platform by discussing the gap between our current capabilities and our vision.

Firstly, the current system does not allow researchers to do self-service exploratory data analysis. Researchers get access to patient data either by filing an IRB request for identified patient data (not exploratory) or via an Honest Broker consultation for anonymized data (not self service). With the growing demand for data, the demand for IRB and Honest Broker services are rising. Growing number of researchers are frustrated with delays in starting their projects due to the arduous IRB process [Silberman2011].  The current system is not designed to support the growing enthusiasm for AI driven research.

We can reduce researcher reliance on Honest Broker services by enabling self-service pre-IRB access to anonymized EHR data. The idea of using anonymized CDW for pre-IRB studies is not new [Liu2009] and exists at other academic centers like UCSF (https://informationcommons.ucsf.edu/). We expect that pre-IRB access to anonymized data will enable research teams to start their studies earlier and publish sooner. For example, with anonymized data a team may realize the need for advanced statistics skillset to account for the inherent biases in EHR data [Verheji2018] much before their IRB request is approved. In some cases, researchers will realize that the underlying data doesn't serve the original study requirements and consequently, they will avoid the arduous IRB process and reduce their cost. Finally, many researchers will be able to complete their research studies with anonymized data and will reduce the number who eventually need access to identified data thus reducing the likelihood of privacy breaches. In summary, enabling pre-IRB access to anonymized EHR data in a self-service fashion will reduce several of our current research bottlenecks and result in faster and better research outcomes.

Secondly, researchers find it difficult to extend their studies on Stanford datasets to non-Stanford data sources. The difficulty arises because we use a data model, STRIDE, that is idiosyncratic and non-standard. To extend the studies to alternate data sources, researchers engage in the complex legal process of data sharing agreements and then spend enormous resources transforming the data from one format to another. We propose to standardize on Observational Health Data Sciences and Informatics (OHDSI) Observational Medical Outcomes Partnership (OMOP) Common Data Model (CDM) [Hripcsak2015], a data model that is widely used by other universities and institutions. The utility of standardization is interoperability [Wilkinson2016] and is demonstrated across hundreds of studies in the clinical data research networks [McCarty2011, Visweswaran2018, Platt2018]. This standardization will mean that Stanford researchers will be able to extend their studies to non-Stanford datasets faster and with ease, while also benefiting from and contributing to the large community of researchers contributing to analytical method development [Vashisht2018, Wang2020].

Our choice of OMOP CDM is motivated by multiple factors. There are a number of popular CDMs to choose from including i2b2 [Murphy2006, Weber2009], Pediatric Learning Healthcare System PEDSNet [Forrest2014], Patient-Centered Clinical Research Network PCORNet [Fleurence2014], Health Care Systems Research Network [Ross2014], and the US Food and Drug Administration Sentinel [Curtis2012]. Choosing a particular CDM over another is a matter of meeting specific research objectives [Kahn2012, Ogunyemi2013, Huser2013, Xu2015]. We choose OHDSI OMOP CDM for its demonstrated applicability for many different use cases including a) claims and EHR [Overhage2012], b) EHR based longitudinal registries [Garza2016] and, c) Hospital transactional database [Makadia2014]. Stanford data in OMOP CDM pilot phase has resulted in several successful studies [Agarwal2016, Banda2016, Hripcsak2016, Duke2017, Vashisht2018]. The OMOP CDM demonstrates strong results in comparative effectiveness research [Ogunyemi2013] with minimal information loss during data transformation [Voss2015], speeds up implementation of clinical phenotypes across networks [Hrispack2019], and promotes research reproducibility [Zhao2018]. There is demonstrated interoperability between different CDMs [Klann2018] so choosing OMOP does not exclude support for other CDMs in future. Furthermore, there is a strong focus in OHDSI community on data quality and broad support for the analytical toolkits (aka methods library) that together strive to deliver consistency in cohort definition, analysis design, and reporting of results. Perhaps the most appealing aspect is that OHDSI is an open source public-private partnership and welcomes community participation. There is a robust community of end users, developers and thought leaders who are actively engaged in various shared repositories, discussion forums, training and workshops. The collection of learning resources are vast (https://github.com/OHDSI/) and includes FAQs, code snippets and video lectures. Finally, OMOP is adopted at other CTSA sites *e.g.*, Albert Einstein College of Medicine – Montefiore Health, Columbia University, Icahn School of Medicine at Mt. Sinai.

Thirdly, current research computing infrastructure is not suited to modern clinical data science. Security issues are significant when handling patient data. It has been shown that even anonymized patient data can be combined with other publicly available data to re-identify patients [Emam2011]. In the hands of a malicious attacker, re-identification can lead to potential harm for vulnerable groups. Clinical notes, even when anonymized, may contain incidental PHI. Researcher laptops, which are secured by Stanford IT, have insufficient storage and compute to support complex analyses on these datasets. Large scale shared computing environments at Stanford, such as Sherlock (https://www.sherlock.stanford.edu/), have the required storage and compute, but are not designed for sensitive patient data. Further, requiring individual labs or researchers to set up servers and manage constantly evolving security issues is infeasible, there are a number of well-publicized significant OS and/or architectural vulnerabilities (https://nvd.nist.gov/) in any year that require sophisticated IT expertise.

We built a shared data science platform that is both secure and can provide the necessary computational power where researchers can run common data science workflows (e.g. TensorFlow using GPGPUs). We believe that such a platform will speed up novel computationally intensive healthcare research such as advanced phenotyping [Deisseroth2018], radiology [Dunnmon2019, Patel2020], echocardiography [Dykes2019], and bedside monitoring [Miller2018]. It will also improve patient data security, and reduce institutional risks.

## Methods:

In this section, we outline the methods used in meeting the three objectives, a) enabling self-service pre-IRB access to anonymized EHR data, b) standardizing the data, and c) supporting researchers on a secure data science platform. Details of these methods are available in the Supplementary section.

Stanford EHR has over a hundred million clinical notes and our goal is to support monthly data refresh. We therefore need a secure infrastructure to support scalable storage and data processing. We expand our HIPAA compliant data center that hosts STRIDE to incorporate a public cloud, Google Cloud Platform (GCP), and use a mix of on-premise private cloud and public cloud capabilities. The hybrid infrastructure meets Stanford's data governance requirements (Supplementary material, Section 1) and minimum security requirements (Supplementary material, Section 2). To support the growing number of researchers using the EHR data, we use cloud native high performance analytical data warehouse Google BigQuery platform (Supplemental material, Section 3) to host and query the data. We adopt open source cloud agnostic workflow runner (Supplementary Material Section 4) for processing data. We anonymize all data including clinical text using best-in-class de-identification methods (Supplementary material, Section 6).

We normalize the EHR data to OMOP CDM v 5.3.1 (Supplementary Material Section 5) in a new analytical warehouse, STARR-OMOP.  The NOTE table in OMOP is used to store the anonymized clinical text. Additionally, we populate the NOTE_NLP table to improve cohort design [Edinger2017]. We use a method that has incorporated both negation detection and history detection to convert clinical terms to known medical concepts (Supplementary material, Section 7).

For a secure data science facility, we collaborate with Stanford's Research Computing (SRC) team, to build a STARR integrated secure data science platform, Nero (Supplementary material, Section 8). Nero is HIPAA compliant and supports Big Data analytics. Armed with modern containerization and container orchestration technologies, the SRC team provides a range of services including hardware procurement and maintenance, OS upgrades, cloud integration, security requirements to meet HIPAA compliance, tool integration and research computing support. Researchers simply need to request access to the secure platform, but otherwise, need not worry about hardware procurement, and system security. In addition to data science tools such as Jupyter notebook, Python, anaconda, TensorFlow, RStudio, SAS etc., the data science platform supports OHDSI analytical tools.

Researchers directly access the anonymized non-human-subject dataset, STARR-OMOP-deid, in the secure Nero platform (Supplementary material, Section 9). The identified STARR-OMOP is available to the Honest Broker team on Nero and the researcher can seek post-IRB OMOP data after IRB approval via the Honest Broker team. Finally, data science training [Dolezel2019] is an important aspect of making the data self service. As a result, with support from Stanford CTSA award, RIT has launched training (Supplementary material, Section 10) on the use of the new anonymized data warehouse.

## Analytics:

In this section, we outline the analytics derived from our new OMOP database and usage for our new data science platform.

We present analytics from our OMOP conversion using NCATS CTSA Common Metrics ((https://ncats.nih.gov/ctsa/projects/common-metrics)) in Table S5 (Supplementary material, Section 5). We show that OMOP database has ~2.7 million patients, over 60% of the patients have a diagnosis (ICD 9/10), over 40% have medication information (RxNorm), ~75% have lab information (LOINC), and over 90% of patients have clinical notes. In figure S5.1 and table S5.2 we show the enhancement in standardized concepts over raw Clarity data.

We present analysis of PHI rate observed in our clinical text processing. Approximately, 100 million clinical notes are processed. These contain 33 billion words, the median number of words per clinical note is ~100 (Figure S5, Supplementary material, Section 5). There are ~22 million notes with no PHI findings, and ~1.3M notes with more than 100 PHI findings. Of the ~33 billion words, 1.4 billion are determined as PHI findings *i.e.*, approximately 4% of the words are found to be PHI (Figure S6.4, Supplementary material, Section 6). We also present the distribution of different PHI types (e.g. name,

MRN, date) found in the notes and success rate of methods in finding these PHI e.g., NLP vs patient look up (Figure S6.5, Supplementary material, Section 6).

We present the number and percent of medical concepts found via text processing in Table S7 (Supplementary material, Section 7). Nearly 2 billion concepts (122K unique concepts) are found in 33 billion words representing 14 different vocabularies.

Finally, we present the performance of Google BigQuery. In Table S3 (Supplementary material, Section 6), we compare Oracle vs BigQuery and show that BigQuery is 10-100x faster. In Table S9.1, we present the aggregate statistical performance of running Achilles Heel, data characterization queries, on STARR-OMOP dataset. Out of a total of 725 queries available in Achilles, an impressive 660 queries took less than 17 seconds, and median execution time was 3 sec. The longest query took 20 min and processed 13 GB of data. We expect that the research community will similarly experience ultra fast performance when using the STARR-OMOP-deid for analysis.

In table S9.3, we present performance of ATLAS running on Google Cloud using Google BigQuery against postgreSQL using the commonly available Medicare Claims Synthetic Public Use datafiles (DE-SynPUF, [https://www.cms.gov/Research-Statistics-Data-and-Systems/Downloadable-Public-Use-Files/SynPUFs/DE_Syn_PUF](https://www.cms.gov/Research-Statistics-Data-and-Systems/Downloadable-Public-Use-Files/SynPUFs/DE_Syn_PUF)). The OHDSI SqlRender package ([https://github.com/OHDSI/SqlRender](https://github.com/OHDSI/SqlRender)) and JDBC drivers require performance tuning to support a new database type, in this case, BigQuery. In nearly all cases, the performance of ATLAS is comparable or faster when using BigQuery.

The Nero platform was launched in the summer of 2018. Nero has on-boarded 400 researchers representing 60 Stanford Principal Investigators. These researchers are working on High Risk data of various types, *i.e.*, not limited to biomedical data. Over 200 TB of storage has already been provisioned and we see a strong growth in use of GPGPUs on-premise. The STARR-OMOP-deid dataset was launched in the fall of 2019. A small subset of our Nero research community, ~30, have completed the user training at the time of writing. The ATLAS ([https://ohdsi-atlas.stanford.edu/](https://ohdsi-atlas.stanford.edu/)) was launched in early 2020.

We propose to present the research impact of the new solutions in a subsequent manuscript. We are also in the process of running network studies using the OMOP database and case studies demonstrating the clinical text processing and will present the results in subsequent manuscripts.

## Future work:

We are bringing in other data types in STARR such as radiology, bedside monitoring and pathology. We propose to extend the methods proposed in meeting the three objectives, a) enabling self-service pre-IRB access to anonymized EHR data, b) standardizing the data, and c) supporting researchers on a secure data science platform, to these other data types.

For example, we can link the datasets first, and then de-identify using the same protocol such that the HIPAA identifiers are deterministic anonymized across all datasets to preserve the linking.

For standardization of other data types, OMOP CDM does not offer a complete solution yet. However, such data can be captured using other well defined standards. For example, radiology Picture Archiving and Communication System (PACS) data is currently in pilot in OMOP CDM (https://github.com/OHDSI/Radiology-CDM), but the underlying DICOM (Digital Imaging and Communications in Medicine, https://www.dicomstandard.org/) data type is a well defined standard for radiology. RIT is building a metadata dataset that contains all metadata from images in the PACS Vendor Neutral Archive. This dataset can be linked to the OMOP dataset via image accession IDs and allow researchers a) unprecedented search capability and, b) run AI models taking the confounding factors such as device models [Badgeley2019] into account. For example, researcher can query the number of studies where patients have neurodegeneration and the image has annotations stored in Grayscale Softcopy Presentation State Storage (GSPS) Service-Object Pair (SOP) class, to estimate the effort in building test sets for their AI model. We have an image de-identification pipeline that is built for petascale data processing and can support both metadata and pixel scrubbing. We propose to present the pipeline in a subsequent manuscript.

Research IT's data center that hosts STARR and related portfolio of data warehouses and tools and data science platform, Nero, are designed for petascale data processing and team science respectively. We expect to integrate with other cloud platforms like AWS and Azure in time.

## Acknowledgment:

STARR suite (2008-), including the first generation data warehouse STRIDE, are made possible by Stanford School of Medicine Dean's Office. User training is supported by the National Center for Research Resources and the National Center for Advancing Translational Sciences, National Institutes of Health, through grant UL1 TR001085.

Using CRediT taxonomy (https://casrai.org/credit/), we present the contributing roles for our authors - Somalee Datta (Conceptualization, Methodology, Supervision, Resources, Writing - original draft), Jose Posada (Data curation, Formal Analysis, Methodology, Validation), Garrick Olson (Supervision, Resources, Investigation) , Wencheng Li (Investigation, Software), Ciaran O'Reilly (Investigation, Software), Deepa Balraj (Data curation, Software), Joe Mesterhazy (Investigation, Software), Joe Pallas (Software), Priyamvada Desai (Conceptualization, Project administration, Validation), Nigam H. Shah (Conceptualization, Writing - review & editing).

For the broader team, we are grateful to Juan Banda (Methodology), Assistant Professor at Georgia State University, who helped us start on the OMOP CDM journey.

We thank Nigam Shah's lab for being early adopters of the dataset, and tools and providing us with feedback and support. We also thank Nivedita Shenoy (Project administration), Agile Program Manager, Technology & Digital Solutions, who helped our engineering team with implementation of the scrum process during the development cycle. We are also thankful to our partners, Stanford Research Computing Center for their collaboration on Nero, Biarca for DevSecOps services, and Odysseus Data Services for OMOP data transformation services. We thank Odysseus Data Services Inc for the performance benchmarking of BigQuery for OHDSI ATLAS tool. We acknowledge the support of the broader Technology & Digital Solutions team, and Michael Halaas (Funding acquisition), Deputy Chief Information Officer of Technology & Digital Solutions and Associate Dean of Industry Relations at Stanford School of Medicine.

## Bibliography:

1. [Agarwal2016] Agarwal V, Podchiyska T, Banda JM, Goel V, Leung TI, Minty EP, Sweeney TE, Gyang E, Shah NH. Learning statistical models of phenotypes using noisy labeled training data. J Am Med Inform Assoc. 2016 May 12. pii: ocw028. doi: 10.1093/jamia/ocw028.
2. [Aronson2010] Aronson AR, Lang F-MM (2010). An overview of MetaMap: historical perspective and recent advances. Journal of the American Medical Informatics Association, 17(3), 229–236. https://doi.org/10.1136/jamia.2009.002733
3. [Badgeley2019] Badgeley MA, Zech JR, Oakden-Rayner L, Glicksberg BS, Liu M, Gale W, McConnell MV, Percha B, Snyder TM & Dudley JT, Deep learning predicts hip fracture using confounding patient and healthcare variables, npj Digital Medicine 2, Article number: 31 (2019)
4. [Banda2016] Banda JM, Callahan A, Winnenburg R, Strasberg HR, Cami A, Reis BY, Vilar S, Hripcsak G, Dumontier M, Shah NH. Feasibility of Prioritizing Drug-Drug-Event Associations Found in Electronic Health Records. Drug Saf. 2016 Jan;39(1):45-57. doi: 10.1007/s40264-015-0352-2.
5. [Banda2019] Banda JM, Sarraju A, Abbasi F, Parizo J, Pariani M, Ison H, Briskin E, Wand H, Dubois S, Jung K, Myers SA, Rader DJ, Leader JB, Murray MF, Myers KD, Wilemon K, Shah NH, Knowles JW, Finding missed cases of familial hypercholesterolemia in health systems using machine learning, npj Digital Medicine volume 2, Article number: 23 (2019)
6. [Banerjee2019] Banerjee I, Sofela M, Yang J, Chen JH, Shah NH, Ball R, Mushlin AI, Desai M, Bledsoe J, Amrhein T, Rubin DL, Zamanian R, Lungren MP, Development and Performance of the Pulmonary Embolism Result Forecast Model (PERFORM) for Computed Tomography Clinical Decision Support, JAMA Netw Open. 2019 Aug 2;2(8):e198719. doi: 10.1001/jamanetworkopen.2019.8719.
7. [Carrell2013] Carrell D, Malin B, Aberdeen J, Bayer S, Clark C, Wellner B, Hirschman L, Hiding in plain sight: use of realistic surrogates to reduce exposure of protected health information in clinical text, J Am Med Inform Assoc. 2013 Mar-Apr; 20(2): 342–348.

8.  [Chapman2001] Chapman WW, Bridewell W, Hanbury P, Cooper GF, Buchanan BG (2001) A simple algorithm for identifying negated findings and diseases in discharge summaries. J Biomed Inform 34:301–310.

9.  [Chapman2007] Chapman WW, Chu D, Dowling JN (2007) ConText: An algorithm for identifying contextual features from clinical text. Proceedings of BioNLP 2007: Biological, translational, and clinical language processing, Association for Computational Linguistics, p. 81-88.

10. [Choi2015]  Choi HJ, Lee MJ, Choi C-M, Lee JH,Shin S-Y, Lyu Y, Park YR, Yoo S, Establishing the role of honest broker: bridging the gap between protecting personal health data and clinical research efficiency, PeerJ. 2015; 3: e1506.

11. [Conway 2019] Conway, M., Keyhani, S., Christensen, L., South, B. R., Vali, M., Walter, L. C., … Chapman, W. W. (2019). Moonstone : a novel natural language processing system for inferring social risk from clinical narratives, 0, 1–10.

12. [Curtis2012] Lesley H. Curtis  Mark G. Weiner  Denise M. Boudreau  William O. Cooper  Gregory W. Daniel  Vinit P. Nair  Marsha A. Raebel  Nicolas U. Beaulieu  Robert Rosofsky  Tiffany S. Woodworth  Jeffrey S. Brown, Design considerations, architecture, and use of the Mini‐Sentinel distributed data system, Pharmacoepidemiology and drug safety2012;21(S1): 23–31

13. [Danciu2014] Danciu I, Cowan JD, Basford M, Wang X, Saip A, Osgood S, Shirey-Rice J, Kirby J, Harris PA, Secondary use of clinical data: The Vanderbilt approach, Journal of Biomedical Informatics, Volume 52, December 2014, Pages 28-35

14. [Datta2016] Datta S, Bettinger K, and Snyder M, Secure cloud computing for genomic data, Nature Biotechnology volume 34, 588–591 (2016). A pre-print of the article is available at https://www.biorxiv.org/content/10.1101/034876v1.full.

15. [Deisseroth2018] Deisseroth CA, Birgmeier J, Bodle EE, Kohler JN, Matalon DR, Nazarenko Y, Genetti CA, Brownstein CA, Schmitz-Abe K, Schoch K, Cope H, Signer R; Undiagnosed Diseases Network, Martinez-Agosto JA, Shashi V, Beggs AH, Wheeler MT, Bernstein JA, and Bejerano G (2018). ClinPhen extracts and prioritizes patient phenotypes directly from medical records to expedite genetic disease diagnosis. Genetics in Medicine, 2018. DOI: 10.1038/s41436-018-0381-1

16. [Deleger2013] Deleger L, Molnar K, Savova G, Xia F, Lingren T, Li Q, Marsolo K, Jegga A, Kaiser M, Stoutenborough L, and Solti I, Large-scale evaluation of automated clinical note de-identification and its impact on information extraction, J Am Med Inform Assoc. 2013 Jan-Feb; 20(1): 84–94.

17. [Dolezel2019] Dolezel D, McLeod A, Big Data Analytics in Healthcare: Investigating the Diffusion of Innovation, Perspect Health Inf Manag, v.16(Summer); PMC 6669368.

18. [Duke2017] Duke JD, Ryan PB, Suchard MA, Hripcsak G, Jin P, Reich C, Schwalm MS, Khoma Y, Wu Y, Xu H, Shah NH, Banda JM, Schuemie MJ, Risk of angioedema associated with levetiracetam compared with phenytoin: Findings of the observational health data sciences and informatics research network, Epilepsia. 2017 Aug;58(8):e101-e106. doi: 10.1111/epi.13828

19. [Dunnmon2019] Dunnmon JA, Yi D, Langlotz CP, Ré C, Rubin DL, Lungren MP. Assessment of Convolutional Neural Networks for Automated Classification of Chest Radiographs. Radiology. 2019 Feb;290(2):537-544. J Vasc Interv Radiol. 2018 Nov;29(11):1527-1534.e1

20. [Dykes2019] Dykes JC, Kipps AK, Chen A, Nourse S, Rosenthal DN, Tierney E, Parental Acquisition of Echocardiographic Images in Pediatric Heart Transplant Patients Using a Handheld Device: A Pilot Telehealth Study, Journal of American Soc. of Echocardiography, 2019; 32 (3): 404–11.

21. [Edinger2017] Edinger T, Demner-Fushman D, Cohen AM, Bedrick S, and Hersh W, Evaluation of Clinical Text Segmentation to Facilitate Cohort Retrieval, AMIA Annu Symp Proc. 2017; 2017: 660–669.

22. [Ellrott2019] Ellrott K, Buchanan A, Creason A, Mason M, Schaffter T, Hoff B, Eddy J, Chilton JM, Yu T, Stuart JM, Saez-Rodriguez J, Stolovitzky G, Boutros PC, Guinney J, Reproducible biomedical benchmarking in the cloud: lessons from crowd-sourced data challenges, Genome Biology volume 20, Article number: 195 (2019)

23. [Emam2011] El Emam K, Jonker E, Arbuckle L, Malin B, A Systematic Review of Re-Identification Attacks on Health Data, PLoS ONE 6(12): e28071. https://doi.org/10.1371/journal.pone.0028071

24. [Evans2012] Evans RS, Lloyd JF, Pierce LA, Clinical Use of an Enterprise Data Warehouse, AMIA Annu Symp Proc. 2012; 2012: 189–198.

25. [Fernandez2013] Ferrández O, South BR, Shen S, Friedlin J, Samore MH, and Meystre SM, BoB, a best-of-breed automated text de-identification system for VHA clinical documents, J Am Med Inform Assoc. 2013 Jan-Feb; 20(1): 77–83.

26. [Fleurence2014] Fleurence RL, Curtis LH, Califf RM, Platt R, Selby JV, Brown JS, Launching PCORnet, a national patient-centered clinical research network, J Am Med Inform Assoc. 2014 Jul-Aug;21(4):578-82. doi: 10.1136/amiajnl-2014-002747.

27. [Foran2017] Foran DJ, Chen W, Chu H,Sadimin E, Loh D,Riedlinger G, Goodell LA, Ganesan S, Hirshfield K,Rodriguez L,DiPaola RS, Roadmap to a Comprehensive Clinical Data Warehouse for Precision Medicine Applications in Oncology, Cancer Inform. 2017; 16: 1176935117694349.

28. [Forrest2014] Forrest C, Margolis P, Bailey L, Marsolo K, Del Beccaro M, Finkelstein J, Milov D, Vieland V, Wolf B, Yu F and Kahn M. PEDSnet: a National Pediatric Learning Health System. J Am Med Inform Assoc, 2014; 21(4): 602-606.

29. [Garza2016 ] Garza M, GuilhermeDel Fiol GD, Tenenbaum J,Walden A, Zozus MN, Evaluating common data models for use with a longitudinal community registry, Journal of Biomedical Informatics, Volume 64, December 2016, Pages 333-341.

30. [Gold2016] Gold M, McLaughlin C, Assessing HITECH Implementation and Lessons: 5 Years Later, Milbank Q. 2016 Sep; 94(3): 654–687.

31. [Goldberger2000] Goldberger AL, Amaral LA, Glass L, Hausdorff JM, Ivanov PC, Mark RG, Mietus JE, Moody GB, Peng CK, Stanley HE, PhysioBank,

PhysioToolkit, and PhysioNet: components of a new research resource for complex physiologic signals, Circulation. 2000 Jun 13;101(23):E215-20.

32. [Gombar2019] Gombar S, Callahan A, Califf R, Harrington R, Shah NH, It is time to learn from patients like mine, npj Digital Medicine volume 2, Article number: 16 (2019).

33. [Grady2015] Grady C, Institutional Review Boards, Purpose and Challenges, Chest. 2015 Nov; 148(5): 1148–1155.

34. [Harkema2009] Harkema, H., Dowling, J. N., Thornblade, T., & Chapman, W. W. (2009). ConText: An algorithm for determining negation, experiencer, and temporal status from clinical reports. Journal of Biomedical Informatics, 42(5), 839–851. https://doi.org/10.1016/j.jbi.2009.05.002

35. [Hernandez-Boussard2016] Hernandez-Boussard T, Tamang S, Blayney D, Brooks J, Shah N, New Paradigms for Patient-Centered Outcomes Research in Electronic Medical Records: An Example of Detecting Urinary Incontinence Following Prostatectomy, EGEMS (Wash DC). 2016 May 12;4(3):1231. doi: 10.13063/2327-9214.1231. eCollection 2016.

36. [Hernandez-Boussard2019] Hernandez-Boussard T, Monda KL, Crespo BC, Riskin D, Real world evidence in cardiovascular medicine: ensuring data validity in electronic health record-based studies, Journal of the American Medical Informatics Association : Jamia, 31 Oct 2019, 26(11):1189-1194

37. [HHS2002] U.S. Department of Health and Human Services (2002) Standards for privacy of individually identifiable health information, final rule, 45 CFR, pt 160–164. U.S. Department of Health and Human Services.

38. [Hripcsak2015] Hripcsak G, Duke JD, Shah NH, Reich CG, Huser V, Schuemie MJ, Suchard MA, Park RW, Wong ICK, Rijnbeek PR, Lei J van der, Pratt N, Norén GK, Li Y-C, Stang PE, Madigan D, and Ryan PB. Observational Health Data Sciences and Informatics (OHDSI): Opportunities for Observational Researchers. Stud Health Technol Inform. 2015; 216: 574–578.

39. [Hripcsak2016] Hripcsak G, Ryan PB, Duke JD, Shah NH, Park RW,  Huser V, Suchard MA, Schuemie MJ, DeFalco FJ, Perotte A, Banda JM, Reich CG, Schilling LM, Matheny ME, Meeker D, Pratt N, Madigan D, Characterizing treatment pathways at scale using the OHDSI network, Proc Natl Acad Sci U S A. 2016 Jul 5; 113(27): 7329–7336.

40. [Hripcsak2019] Hripcsak G, Shang N, Peissig PL, Rasmussen LV, Liu C, Benoit B, Carroll RJ, Carrell DS, Denny JC, Dikilitas O, Gainer VS, Howell KM, Klann JG, Kullo IJ, Lingren T, Mentch FD, Murphy SN, Natarajan K, Pacheco JA, Wei WQ, Wiley K, Weng C, Facilitating phenotype transfer using a common data model. J Biomed Inform. 2019 Aug;96:103253. doi: 10.1016/j.jbi.2019.103253..

41. [Huser2013] Huser V, Cimino JJ, Desiderata for Healthcare Integrated Data Repositories Based on Architectural Comparison of Three Public Repositories, AMIA Annu Symp Proc. 2013; 2013: 648–656.

42. [Johnson2016] Johnson AEW, Pollard TJ, Shen L, Lehman L-WH, Feng M, Ghassemi M, Moody B, Szolovits P, Celi LA,  Mark RG, MIMIC-III, a freely accessible critical care database, Scientific Data 3, Article number: 160035 (2016).

43. [Jung2019] Jung K, Sudat SEK, Kwon N, Stewart WF, Shah NH, Predicting need for advanced illness or palliative care in a primary care population using electronic health record data, J Biomed Inform. 2019 Apr;92:103115. doi: 10.1016/j.jbi.2019.103115. Epub 2019 Feb 10.

44. [Kahn2012] Kahn MG, Batson D, Schilling LM, Data Model Considerations for Clinical Effectiveness Researchers, Med Care. 2012 Jul; 50(0): doi: 10.1097/MLR.0b013e318259bff4

45. [Klann2018] Klann JG, Phillips LC, Herrick C, Joss MAH, Wagholikar KB, Murphy SN, Web services for data warehouses: OMOP and PCORnet on i2b2, Journal of the American Medical Informatics Association, Volume 25, Issue 10, October 2018, Pages 1331–1338, https://doi.org/10.1093/jamia/ocy093

46. [Kunjan2015] Kunjan K, Toscos T, Turkcan A, Doebbeling BN, A Multidimensional Data Warehouse for Community Health Centers, AMIA Annu Symp Proc. 2015; 2015: 1976–1984.

47. [Kurtzer2017] Singularity: Scientific containers for mobility of compute, Kurtzer GM, Sochat V, Bauer MW, PLoS ONE, 12(5), May 11 2017, https://doi.org/10.1371/journal.pone.0177459

48. [Kulikowski2019] Kulikowski CA, Beginnings of Artificial Intelligence in Medicine (AIM): Computational Artifice Assisting Scientific Inquiry and Clinical Art – with Reflections on Present AIM Challenges, Yearb Med Inform. 2019 Aug; 28(1): 249–256.

49. [Leipzig2017] Leipzig J., A review of bioinformatic pipeline frameworks, Brief Bioinform. 2017 May 1;18(3):530-536. doi: 10.1093/bib/bbw020.

50. [LePendu2012] Lependu, P., Iyer, S. V, Fairon, C., & Shah, N. H. (2012). Annotation Analysis for Testing Drug Safety Signals using Unstructured Clinical Notes. Journal of Biomedical Semantics, 3 Suppl 1, S5. https://doi.org/10.1186/2041-1480-3-S1-S5

51. [LePendu2013] LePendu P, Iyer SV, Bauer-Mehren A, Harpaz R, Mortensen JM, Podchiyska T, Ferris TA, Shah NH, Pharmacovigilance using clinical notes, Clin Pharmacol Ther. 2013 Jun;93(6):547-55. doi: 10.1038/clpt.2013.47. Epub 2013 Mar 4.

52. [Levin2019] Levin MA, Lin HM, Prabhakar G, McCormick PJ, Egorova NN, Alive or dead: Validity of the Social Security Administration Death Master File after 2011, Health Serv Res. 2019 Feb;54(1):24-33. doi: 10.1111/1475-6773.13069

53. [Liu2009] Liu J, Erdal S, Silvey SA,  Ding J, Riedel JD, Marsh CB, Kamal J, Toward a Fully De-identified Biomedical Information Warehouse, AMIA Annu Symp Proc. 2009; 2009: 370–374.

54. [Lowe2009] Lowe HJ, Ferris TA, Hernandez PM, Weber SC, STRIDE - An Integrated Standards-Based Translational Research Informatics Platform, AMIA Annu Symp Proc. 2009 Nov 14;2009:391-5.

55. [Maeda2012] Maeda K, Comparative survey of object serialization techniques and the programming supports, J. Commun. Comput., 9 (2012), pp. 920-928

56. [Makadia2014] Makadia R, Ryan PB, Transforming the Premier Perspective® Hospital Database into the Observational Medical Outcomes Partnership (OMOP) Common Data Model, EGEMS (Wash DC). 2014; 2(1): 1110, Published online 2014 Nov 11. doi: 10.13063/2327-9214.1110

57. [Manning2014] Manning CD, Surdeanu M, Bauer J, Finkel J, Bethard SJ, and McClosky D. 2014. The Stanford CoreNLP Natural Language Processing Toolkit In Proceedings of the 52nd Annual Meeting of the Association for Computational Linguistics: System Demonstrations, pp. 55-60.

58. [McCarty2011] McCarty CA, Chisholm RL, Chute CG, Kullo IJ, Jarvik GP, Larson EB, Li R, Masys DR, Ritchie MD, Roden DM, Struewing JP, Wolf WA; eMERGE Team, The eMERGE Network: a consortium of biorepositories linked to electronic medical records data for conducting genomic studies, BMC Med Genomics. 2011 Jan 26;4:13. doi: 10.1186/1755-8794-4-13.

59. [McPadden2019] McPadden J, Durant TJS, Bunch DR, Coppi A, Price N, Rodgerson K, Torre CJ, Byron W, Hsiao AL, Krumholz HM, and Schulz WL, Health Care and Precision Medicine Research: Analysis of a Scalable Data Science Platform, J Med Internet Res. 2019 Apr; 21(4): e13043.

60. [Miller2018] Miller D, Ward A, Bambos N, Scheinker D and Shin A, Physiological Waveform Imputation of Missing Data using Convolutional Autoencoders. In 2018 IEEE 20th International Conference on e-Health Networking, Applications and Services (Healthcom) (pp. 1-6). IEEE. 2018, September.

61. [Munafò2017] Munafò MR, Nosek BA, Bishop DVM, Button KS, Chambers CD, Percie du Sert N, Simonsohn U, Wagenmakers E-J, Ware JJ & Ioannidis JPA, A manifesto for reproducible science, Nature Human Behaviour volume 1, Article number: 0021 (2017)

62. [Murphy2006] Murphy SN, Mendis ME, Berkowitz DA, Kohane I, Chueh HC. Integration of clinical and genetic data in the i2b2 architecture. AMIA Annu Symp Proc. 2006:1040

63. [NISTIR8053] De-Identification of Personal Information, https://nvlpubs.nist.gov/nistpubs/ir/2015/NIST.IR.8053.pdf, http://dx.doi.org/10.6028/NIST.IR.8053

64. [NISTSP800-190] Application Container Security Guide, https://nvlpubs.nist.gov/nistpubs/SpecialPublications/NIST.SP.800-190.pdf, https://doi.org/10.6028/NIST.SP.800-190

65. [Pan2017] Pan C, McInnes G, Deflaux N, Snyder S, Bingham J, Datta S, Tsao PS, Cloud-based interactive analytics for terabytes of genomic variants data, Bioinformatics. 2017 Dec 1; 33(23): 3709–3715.

66. [Parthipan2019] Parthipan, A., Banerjee, I., Humphreys, K., Asch, S. M., Curtin, C., Carroll, I., Hernandez-Boussard, T., Predicting inadequate postoperative pain management in depressed patients: A machine learning approach. PloS one 2019; 14 (2): e0210575

67. [Patel2020] Patel BN, Boltyenkov AT, Martinez, MG, Mastrodicasa D, Marin D, Brooke JR, Chung B, Pandharipande P, Kambadakone A, Cost-effectiveness of dual-energy CT versus multiphasic single-energy CT and MRI for characterization of incidental indeterminate renal lesions

68. [Platt2018] Platt R, Brown JS, Robb M, McClellan M, Ball R, Nguyen MD, Sherman RE, The FDA Sentinel Initiative — An Evolving National Resource, N Engl J Med 2018; 379:2091-2093, DOI: 10.1056/NEJMp1809643

69. [Ogunyemi2013] Ogunyemi O, Meeker D, Kim HE, Ashish N, Farzaneh S, Boxwala A, Identifying appropriate reference data models for comparative effectiveness research (CER) studies based on data from clinical information systems,  Med Care. 2013 Aug;51(8 Suppl 3):S45-52. doi: 10.1097/MLR.0b013e31829b1e0b.

70. [Overhage2012] Overhage JM, Ryan PB, Reich CG, Hartzema AG, Stang PE, Validation of a common data model for active safety surveillance research, J Am Med Inform Assoc. 2012 Jan-Feb;19(1):54-60. doi: 10.1136/amiajnl-2011-000376. Epub 2011 Oct 28.

71. Rajkomar2019] Rajkomar A, Dean J, Kohane I, Machine Learning in Medicine, N Engl J Med. 2019 Apr 4;380(14):1347-1358. doi: 10.1056/NEJMra1814259.

72. [Rajpurkar2018] Rajpurkar P, Irvin J, Ball RL, Zhu K, Yang B, Mehta H, Duan T, Ding D, Bagul A, Langlotz CP, Patel BN, Yeom KW, Shpanskaya K, Blankenberg FG, Seekins J, Amrhein TJ, Mong DA, Halabi SS, Zucker EJ, Ng AY, Lungren MP. Deep learning for chest radiograph diagnosis: A retrospective comparison of the CheXNeXt algorithm to practicing radiologists. PLoS Med. 2018 Nov 20;15(11):e1002686.

73. [Ross2014] Ross TR, Ng D, Brown JS, The HMO research network virtual data warehouse: a public data model to support collaboration, EGEMS, 2 (1) (2014), p. 1049, doi: 10.1306372327-9214.1049.

74. [Ross2019] Ross EG, Jung K, Dudley JT, Li L, Leeper NJ, Shah NH, Predicting Future Cardiovascular Events in Patients With Peripheral Artery Disease Using Electronic Health Record Data, Circulation: Cardiovascular Quality and Outcomes. 2019;12:e004741, https://doi.org/10.1161/CIRCOUTCOMES.118.004741

75. [RRF2009] UMLS® Reference Manual [Internet]. Bethesda (MD): National Library of Medicine (US); 2009 Sep-. 3, Metathesaurus - Rich Release Format (RRF) Available from: https://www.ncbi.nlm.nih.gov/books/NBK9685/

76. [Savova2010] Savova, G. K., Masanz, J. J., Ogren, P. V, Zheng, J., Sohn, S., Kipper-Schuler, K. C., & Chute, C. G. (2010). Mayo clinical Text Analysis and Knowledge Extraction System (cTAKES): architecture, component evaluation and applications. Journal of the American Medical Informatics Association, 17(5), 507–513. Retrieved from http://jamia.oxfordjournals.org/content/17/5/507.abstract

77. [Shah2019] Shah NH, Milstein A, Bagley PhD SC, Making Machine Learning Models Clinically Useful. JAMA. 2019 Aug 8. doi: 10.1001/jama.2019.10306.

78. [Silberman2011] Silberman G and Kahn KL, Burdens on Research Imposed by Institutional Review Boards: The State of the Evidence and Its Implications for Regulatory Reform, Milbank Q. 2011 Dec; 89(4): 599–627.

79. [Stubbs2014] Stubbs A, Kotfila C, Uzunerb Ö, Automated systems for the de-identification of longitudinal clinical narratives: Overview of 2014

i2b2/UTHealth shared task Track 1, Journal of Biomedical Informatics, Volume 58, Supplement, December 2015, Pages S11-S19.

80. [Tamang2015] Tamang S, Patel MI, Blayney DW, Kuznetsov J, Finlayson SG, Vetteth Y, Shah N, Detecting unplanned care from clinician notes in electronic health records, J Oncol Pract. 2015 May;11(3):e313-9. doi: 10.1200/JOP.2014.002741.

81. [Tamang2017a] Tamang SR, Hernandez-Boussard T, Ross EG, Gaskin G, Patel MI, Shah NH, Enhanced Quality Measurement Event Detection: An Application to Physician Reporting, Egems (Washington, DC), 29 May 2017, 5(1):5

82. [Tamang2017b] Tamang S1, Milstein A1, Sørensen HT2, Pedersen L2, Mackey L3, Betterton JR3, Janson L3, Shah N1., Predicting patient 'cost blooms' in Denmark: a longitudinal population-based study, BMJ Open. 2017 Jan 11;7(1):e011580. doi: 10.1136/bmjopen-2016-011580.

83. [Tseytlin2016] Tseytlin, E, Mitchell, K, Legowski, E., Corrigan, J., Chavan, G., & Jacobson, R. S. (2016). NOBLE - Flexible concept recognition for large-scale biomedical natural language processing. BMC Bioinformatics, 17(1), 32. https://doi.org/10.1186/s12859-015-0871-y

84. [Turley2016] Turley CB, Obeid J, Larsen R, Fryar KM, Lenert L, Bjorn A, Lyons G, Moskowitz J, Sanderson I, Leveraging a Statewide Clinical Data Warehouse to Expand Boundaries of the Learning Health System, EGEMS (Wash DC). 2016 Dec 6;4(1):1245. doi: 10.13063/2327-9214.1245.

85. [UMLS2009] UMLS® Reference Manual [Internet]. Bethesda (MD): National Library of Medicine (US); 2009 Sep-. 1, Introduction to the UMLS (https://www.ncbi.nlm.nih.gov/books/NBK9675/)

86. [Vashisht2018] Vashisht R, Jung K, Schuler A, Banda JM, Park RW, Jin S, Li L, Dudley JT, Johnson KW, Shervey MM, Xu H, Wu Y, Natrajan K, Hripcsak G, Jin P, Van Zandt M, Reckard A, Reich CG, Weaver J, Schuemie MJ, Ryan PB, Callahan A, Shah NH. Association of Hemoglobin A1c Levels With Use of Sulfonylureas, Dipeptidyl Peptidase 4 Inhibitors, and Thiazolidinediones in Patients With Type 2 Diabetes Treated With Metformin: Analysis From the Observational Health Data Sciences and Informatics Initiative. JAMA Network Open 2018, 1(4), pp.e181755-e181755.

87. [Verheji2018] Verheij RA, Curcin V, Delaney BC, McGilchrist MM, Possible Sources of Bias in Primary Care Electronic Health Record Data Use and Reuse, J Med Internet Res. 2018 May; 20(5): e185, doi: 10.2196/jmir.9134.

88. [Visweswaran2018] Visweswaran S, Becich MJ, D'Itri VS, Sendro ER, MacFadden D, Anderson NR, Allen KA, Ranganathan D, Murphy SN, Morrato EH, Pincus HA, Toto R, Firestein GS, Nadler LM, Reis SE, Accrual to Clinical Trials (ACT): A Clinical and Translational Science Award Consortium Network, JAMIA Open, Volume 1, Issue 2, October 2018, Pages 147–152, https://doi.org/10.1093/jamiaopen/ooy033

89. [Vivian2017] Vivian J, Rao AA, Nothaft FA, Ketchum C, Armstrong J, Novak A, Pfeil J, Narkizian J, Deran AD, Musselman-Brown A, Schmidt H, Amstutz P, Craft B, Goldman M, Rosenbloom K, Cline M, O'Connor B, Hanna M, Birger C, Kent WJ, Patterson DA, Joseph AD, Zhu J, Zaranek S, Getz G, Haussler D & Paten P,

Toil enables reproducible, open source, big biomedical data analyses, Nature Biotechnology volume 35, pages314–316(2017).

90. [Voss2015] Voss EA, Makadia R, Matcho A, Ma Q, Knoll C, Schuemie M, DeFalco FJ, Londhe A, Zhu V, Ryan PB, Feasibility and utility of applications of the common data model to multiple, disparate observational health databases, J Am Med Inform Assoc. 2015 May; 22(3): 553–564.

91. [Wang2018] Wang JK, Hom J, Balasubramanian S, Schuler A, Shah NH, Goldstein MK, Baiocchi MTM, Chen JH, An Evaluation of Clinical Order Patterns Machine-Learned from Clinician Cohorts Stratified by Patient Mortality Outcomes, J Biomed Inform. 2018 Oct;86:109-119. doi: 10.1016/j.jbi.2018.09.005.

92. [Wang2020] Wang Q, Reps JM, Kostka KF, Ryan PB, Zou Y, Voss EA, Rijnbeek PR, Chen R, Rao GA, Stewart MH, Williams AE, Williams RD, Van Zandt M, Falconer T, Fernandez-Chas M, Vashisht R, Pfohl SR, Shah NH, Kasthurirathne SN, You SC, Jiang Q, Reich C, Zhou Y, Development and validation of a prognostic model predicting symptomatic hemorrhagic transformation in acute ischemic stroke at scale in the OHDSI network, PLoS One. 2020 Jan 7;15(1):e0226718. doi: 10.1371/journal.pone.0226718. eCollection 2020.

93. [Weber2009] Weber GM, Murphy SN, McMurry AJ, Macfadden D, Nigrin DJ, Churchill S, Kohane IS, The Shared Health Research Information Network (SHRINE): a prototype federated query tool for clinical data repositories. J Am Med Inform Assoc. 2009 Sep-Oct;16(5):624-30. PMC2744712.

94. [Weber2010] Weber SC, Lowe HJ, Malunjkar S, Quinn J, Implementing a Real-time Complex Event Stream Processing System to Help Identify Potential Participants in Clinical and Translational Research Studies., AMIA Annu Symp Proc. 2010 Nov 13;2010:472-6

95. [Wilkinson2016] Wilkinson MD, Dumontier M, …, Mons B,The FAIR Guiding Principles for scientific data management and stewardship, Scientific data 3 (2016).

96. [Wilson2017] Wilson S, Fitzsimons M, Ferguson M, Heath A, Jensen M, Miller J, Murphy MW, Porter J, Sahni H, Staudt L, Tang Y, Wang Z, Yu C, Zhang J, Ferretti V, Grossman RL; GDC Project, Developing Cancer Informatics Applications and Tools Using the NCI Genomic Data Commons API, Cancer Res. 2017 Nov 1;77(21):e15-e18. doi: 10.1158/0008-5472.CAN-17-0598.

97. [Xu2010] Xu R, Musen MA, Shah NH (2010) A Comprehensive Analysis of Five Million UMLS Metathesaurus Terms Using Eighteen Million MEDLINE Citations. AMIA Annu Symp Proc 2010:907–911

98. [Xu2015] Xu Y, Zhou X, Suehs BT, Hartzema AG, Kahn MG, Moride Y, Sauer BC, Liu Q, Moll K, Pasquale MK, Nair VP, Bate A, A Comparative Assessment of Observational Medical Outcomes Partnership and Mini-Sentinel Common Data Models and Analytics: Implications for Active Drug Safety Surveillance, Drug Safety Volume 38, Issue 8, 28 August 2015, Pages 749-765

99. [Zhao2018] Zhao Y, Wang Y, Wang H, Yan B, Shen F, Peterson KJ, Walter A, Annotating Cohort Data Elements with OHDSI Common Data Model to Promote

Research Reproducibility, 2018 IEEE International Conference on Bioinformatics and Biomedicine (BIBM), https://ieeexplore.ieee.org/document/8621269

# Supplementary Material

In the supplementary material, we present details of data governance, data security implementation, technological and scientific methodologies used to generate the anonymized and standardized database, the secure data science environment and user support.

## Section 1: Data governance policies

In this section, we present the data governance requirements for access to patient data, both anonymized and identified, and our workflow to meet the requirements.

The raw data in STARR data lake is Protected Health Information and is governed by the standards for privacy of individually identifiable health information by U.S. Department of Health and Human Services [HHS2002]. Research access to identified data (e.g., the identified STARR-OMOP) requires the researcher to get IRB approval specific to their study.

The requirements for the security and privacy of health information established by the Health Insurance Portability and Accountability Act (HIPAA) Privacy Rule and HITECH Act explicitly do not apply to the protected health information (PHI) that has been de-identified, provided that there is "no reasonable basis to believe that the information can be used to identify an individual". Such de-identified or limited data sets do not require IRB approval, only Privacy Board Review and an appropriate Data Use Agreement.

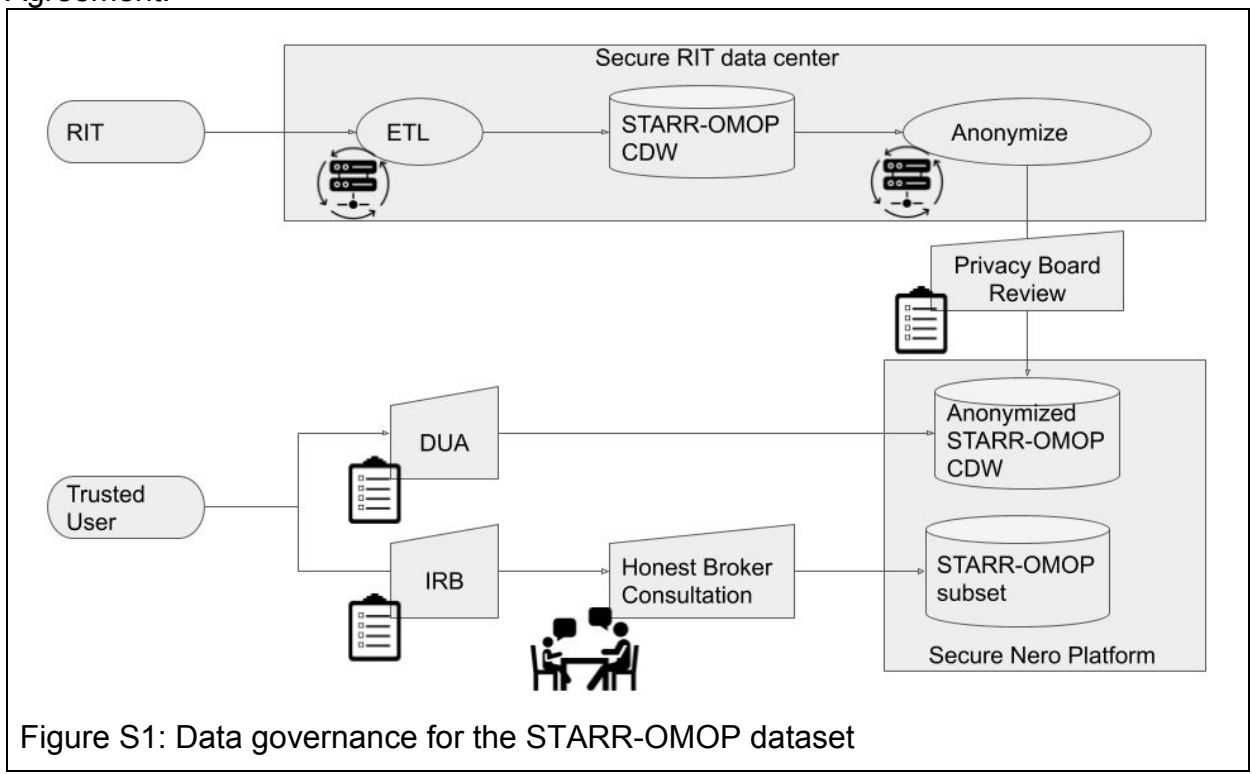

Figure S1: Data governance for the STARR-OMOP dataset

The workflow is illustrated in Figure S1. Research IT de-identifies the patient data (Section 6), in STARR-OMOP, into an anonymized dataset STARR-OMOP-deid and undertakes the Privacy Board Review. For the STARR-OMOP-deid dataset to be classified as non human subject research, Research IT does not allow subsequent linking the de-identified person IDs in the anonymized dataset to patient MRNs.

Researchers as part of STARR-OMOP-deid data access, sign a DUA referred to as a Data Privacy Attestation that prohibits recipients of anonymized data from attempting to re-identify the data subjects, or from resharing the data without permission.

Patient data anonymization is required but is not sufficient to maintain privacy. It has been shown that de-identified data can be combined with other publicly available data to re-identify patients [Emam2011]. If the de-identified data is put on servers that are subsequently compromised, the data can make its way to a malicious privacy hacker and re-identification can lead to potential harm for vulnerable groups such as patients who may face social repercussions due to their LGBQ status, substance abuse, sexually transmitted diseases, suicidal ideations, or mental health issues. Therefore, for data security, STARR-OMOP-deid data is only available in a highly secure computing and storage environment, Nero (Section 8).

## Section 2: Security management

In this section, we present the data security requirements and our security methods.

Stanford University Information Security Office (ISO) provides risk classification for various data types (https://uit.stanford.edu/guide/riskclassifications). The anonymized STARR-OMOP-deid is classified as High Risk data, we believe that loss of confidentiality, integrity, or availability of such data could have a significant adverse impact on our mission, reputation and safety. The technical reason for the High Risk classification is the potential of incidental PHI in the de-identified clinical text. However, as mentioned previously,  de-identified clinical data carries re-identification risk [Emam2011], and safeguarding this data to strictest standards minimizes such re-identification risks. Stanford ISO also provides minimum security guidelines (https://uit.stanford.edu/guide/securitystandards) for different data risk classifications. RIT's infrastructure is required to meet the minimum security guidelines for HIPAA and High Risk data.

The minimum security standards for High Risk data lead to complex set of IT requirements for servers and laptops that access such data.  For example, endpoints such as laptops used by researchers to view data, need to be encrypted, backed up, inventoried, malware protected and patched regularly. Stanford laptops are configured and managed to meet these requirements and prior to CDW data access, user is required to complete an "AM I Encrypted (aka amie)" attestation that is monitored by Stanford School of Medicine. Servers additionally require firewall protection, intrusion detection, centralized logging among other requirements. All systems are required to support two-factor authentication and Research IT additionally supports Stanford's two factor authentication. All systems with High Risk data are required to support audit capabilities.

For scalability, we expanded our on-premise data center to cloud. Google Cloud Platform (GCP) is our first cloud based data center. Under Stanford's business arrangements, while GCP hosts Stanford data, Google does not have access to Stanford data. Infrastructure-as-a-Service (IaaS) and Software-as-a-Service (SaaS) providers such as GCP are required to be covered under a Business Associate Agreement (BAA) if the data resides on provider platforms. However, a BAA is a necessary but not sufficient condition for data security. Cloud services such as GCP or Amazon Web Services (AWS) used by RIT need to be secured by RIT to meet ISO security guidelines for minimum security including guidelines for ephemeral services, containerized solutions, credential and key management. We only use cloud services that are covered under Stanford BAA. For example, for Google Cloud Platform, we use the HIPAA compliant services covered under Stanford-Google BAA (https://cloud.google.com/security/compliance/hipaa-compliance/). Our approach to securing a Cloud IaaS provider is further described in a prior manuscript [Datta2016].

Both the STARR and Nero data science platform use containers heavily. Containers allow a developer to package up an application with all of the parts it needs, such as libraries and other dependencies, and ship it all out as one package. This ability not only streamlines software deployment and security management [NISTSP800-190], it improves user experience and results in a reproducible analytics environment.  We use both Docker (https://www.docker.com/why-docker) and Singularity (https://sylabs.io/singularity/). Docker is used  for Cloud and on-premise infrastructure management. Singularity is a secure container technology that is particularly well suited for researcher (non system admin role) use in HPC environments. Stanford has been one of the early adopters of Singularity for HPC clusters [Kurtzer2017] and as a result, the Stanford Research Computing Center (SRCC) team has tremendous experience supporting our scientific research community using containerized application bundles.

While minimum security requirements may be clear, their implementation is often complex. For the on-premise servers, OS vulnerability management is a constant effort. Cloud providers take on the burden of maintaining secure OS images. On cloud, the complexity often comes from our lack of control over cloud products. For example, cloud providers may introduce features that unwittingly encourage users to export or use data in a non-approved or non-compliant manner. We may not be able to disable those features in a meaningful manner, and attempting to train users not to use those features may be ineffective. Furthermore, the default permission models do not distinguish very effectively between project resource visibility (configuration, metadata, resources, billing, etc.) and High Risk data visibility. It is often hard and complicated to provide effective separation of duties. Finally, the Cloud services are evolving over time, and the various products are not necessarily compatible with each other at a given point in time.

For example, Healthcare APIs and Cloud SQL import/export don't work with VPC Service Controls at the time of writing.

Both STARR and Nero go through Stanford Data Risk Assessment (DRA) process (https://uit.stanford.edu/security/dra) overseen by ISO and UPO. The DRA process is required at Stanford for new High Risk infrastructure or significant changes to existing infrastructure. These platforms also go through external penetration testing from time to time and ISO is part of the team that selects the pentest service provider, reviews the testing protocol and the final report. Prior to the launch of ATLAS Cohort tool (https://ohdsi-atlas.stanford.edu/), we completed an external penetration testing and addressed the vulnerabilities in collaboration with ATLAS development team.

## Section 3: Database technologies

In this section, we describe various database technologies used in our warehouses and provide justification for selection.

Our workflows move data from on-premise relational databases (e.g. Microsoft MSSQL) to cloud native distributed databases (e.g. Google Cloud BigQuery) via Apache Avro (https://avro.apache.org/) format. Avro provides interoperability for databases at rest. Avro is an open, compact binary data serialization format for data intensive application with wide vendor support. Avro files are self-describing (they include the schema) and are the preferred file storage format for many popular Big Data systems such as Apache Hadoop (https://hadoop.apache.org/), Google BigQuery, and Amazon Redshift (https://aws.amazon.com/redshift/). Avro has support for a variety of programming languages such as Java, Scala, C, C++ and Python [Maeda2012]. Avro uses JSON as an explicit schema or dynamically generates schemas of the existing Java objects. Since data is stored with its schema (self-describing), Avro is compatible with scripting languages. While not perfect (for example Avro currently does not have a way to represent a civil date-time) these exceptions can generally be worked around. We receive the Hospital data from a database snapshot. Apache Sqoop (https://sqoop.apache.org/) and Apache Nifi (https://nifi.apache.org/) are the most popular open-source software solutions for exporting databases to Avro format.

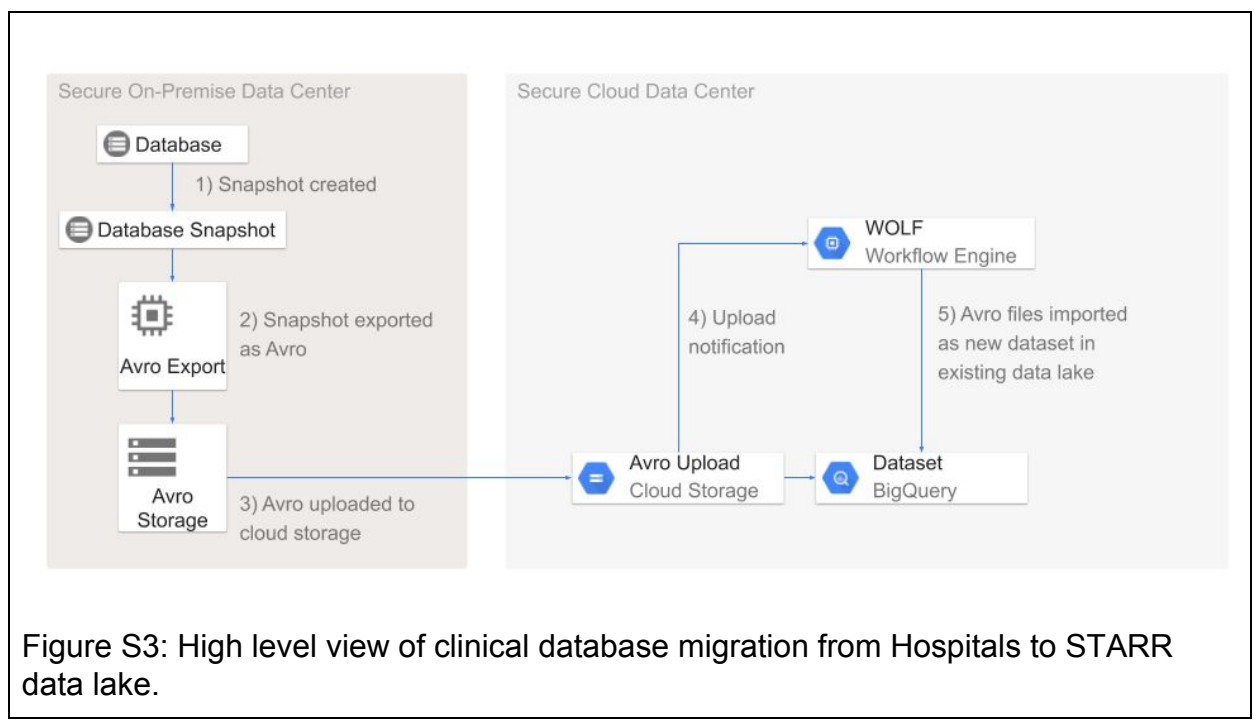

Figure S3: High level view of clinical database migration from Hospitals to STARR data lake.

Once in Avro, the databases are then uploaded to cloud and imported in GCP BigQuery (BQ, https://cloud.google.com/bigquery/) as a dataset for subsequent processing (see Figure S3). BQ is a highly performant analytical data warehousing technology that supports ANSI-compliant SQL and a powerful Application Programming Interface (API). It is not a transactional database like postgreSQL. While Stanford has exceptional on-premise computing resources, we believe that BQ provides a better overall query performance experience with significantly less IT overhead. It is not a low latency workflow, but the database scales exceptionally well for Big Data. BQ has proven to be highly performant for large scale GWAS [Pan2017].

At Stanford, we benchmarked a set of five complex queries using our highly optimized Oracle database running on the state-of-the-art on-premise infrastructure, and the same data, as BigQuery dataset via Avro (Figure S3). These queries were submitted by the Honest Broker team as representative as typical queries requiring complex joins. The results of the benchmarking are presented in Table S3. We see that queries on BigQuery execute 10-100x faster than on Oracle.

| # | Time on BigQuery (in sec) | Time on Oracle (in sec) | Processed data size (in GB) |
|---|---|---|---|
| 1 | 3.2 | 521 | 2.5 |
| 2 | 6.3 | 704 | 19.8 |
| 3 | 47 | 867 | 152.9 |

| 4 | 9.03 | 587 | 24.5 |
| 5 | 102 | 1,200 | 33.3 |

Table S3: Performance benchmarking of Oracle vs BigQuery, using a set of five complex queries.

# Section 4: Data processing

In this section, we present our strategy for reproducible cloud agnostic petascale data processing using open standards.

We leverage the significant investment in development and evaluation of workflow orchestration frameworks (aka workflow engines or pipeline frameworks) [Leipzig2017] in recent years. For our workflow framework (Figure S4), we adopt Common Workflow Language (CWL, https://www.commonwl.org/) and Cromwell workflow execution engine (https://github.com/broadinstitute/cromwell). We evaluated a number of existing workflow orchestration frameworks by deploying and testing the software - Apache Airflow (https://airflow.apache.org/), Cromwell, Toil [Vivian2017], and Arvados (https://arvados.org/). Cromwell workflow executor engine was selected due to its broad range of platform support, its maturity on the GCP, and its ease of deployment.

To support downstream reproducible science [Munafò2017], and data management efficiency, we consider the following requirements to be critical for our data processing:
- Ability to track provenance of all outputs and every intermediate step - track every dependent source data and intermediate transformation so consumer of output has full visibility and understanding of where it came from. In our design, the source data is immutable (snapshots, not live databases or streams).
- Ability to control reproducibility of all outputs including, a) ability to archive all versions of inputs including option to support purging to manage cost vs. reproducibility tradeoff, b) ability to transform in a deterministic fashion, c) ability to retain all software or logic used for transformation steps so we can run it in the future.
- Ability to efficiently re-run transformations when new versions (snapshots) of the source data become available *i.e.*, run transformations when things change, not on a schedule
- Ability to add quality checks in each transformation step, so pipelines fail early rather than letting errors propagate

Aside from selection of a workflow manager, these requirements also lead to design considerations. For example, is a monolithic step in the transformation pipeline better than loosely coupled steps? Loosely coupled steps are better for progress monitoring, workflow transparency, and enable intermediate stop and restart while tightly coupled steps offer better reproducibility and version compatibility. In another example, what should constitute the archival "raw" data? While the original database (e.g., Oracle) dump can be stored as the raw, it is a proprietary and large file. Alternatively, the highly compressed and standard avro can be the "raw" but is also less reliable due to potential bugs in Oracle technology stack or our data processing code used transformation steps.

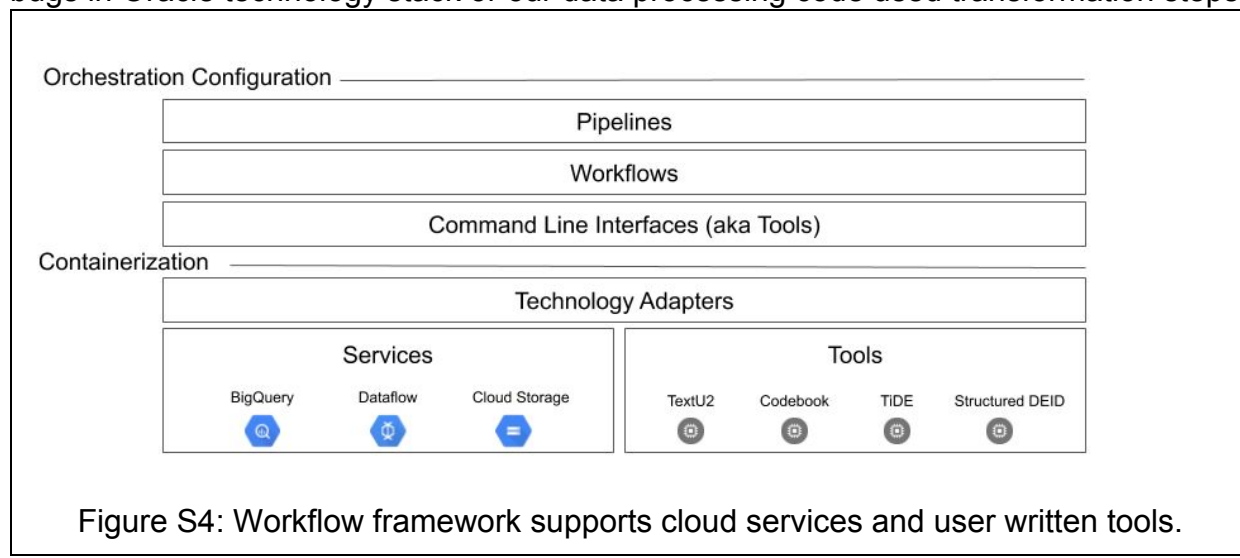

Figure S4: Workflow framework supports cloud services and user written tools.

A STARR data pipeline corresponds to a top level CWL workflow. Configuration is the CWL definitions of workflows and their command line interfaces (Tools). Technology adapters are implemented through the use of containerization (Docker) to better support reproducibility and portability. These adapters are organized into two main categories:

A. Tool adapters consisting of self contained containers that focus on a specific set of functionality, for example, text de-identification. Scaling of tools is explicitly managed by the workflow framework by instantiating multiple instances across a distributed computing resource along with managing the partitioning of work among instances.

B. Service adapters to general purpose technologies that reside outside the containers in which the adapters live. Services are self scaling and not dependent on the workflow framework for scalability. For example, for BigQuery, our workflow framework provides a service adapter with similar capabilities to Google's BigQuery command line tool (https://cloud.google.com/bigquery/docs/reference/bq-cli-reference). However, instead of simply wrapping this command line tool, its capabilities are mimicked where appropriate by leveraging BigQuery Java Client APIs (https://googleapis.dev/java/google-cloud-clients/latest/index.html?com/google/cloud/bigquery/package-summary.html) so that functionality can be more easily adapted and extended to work within the framework. Additions in functionality include, i) Data-quality-checks that enables the execution of a BigQuery measurement query and then the execution of a list of predefined rules to ensure the measurements obtained match a given set of criteria (for example no null values in a column), ii) View-dataset that enables the creation/update of a dataset that comprises of authorized views on another dataset (for example, maintain a statically named latest dataset of a series of snapshot datasets), and

iii) Metadata that adds support for querying the tables in a dataset and individual table schemas in a way that interoperates with CWL's command line interface.

# Section 5: Data transformation from EHR to OMOP CDM

In this section, we highlight the ETL (Extract, Transform, Load) processes that take us from Clarity data to OMOP data.

RIT receives Epic Clarity Electronic Health Records (EHR) data from the two Stanford Hospitals: Adult Hospital (aka Stanford Health Care, https://stanfordhealthcare.org/) and Children's Hospital (aka Lucile Packard Children's Hospital, https://www.stanfordchildrens.org/). Epic Clarity contains data from the clinics that are part of the University Healthcare Alliance (https://universityhealthcarealliance.org/) and Lucile Packard Healthcare Alliance (https://www.stanfordchildrens.org/en/about/our-network). RIT also receives HL7 feeds from the two Hospitals such as Admit Discharge Transfer (ADT), and Billing Account Record (BAR) and the Social Security Administration Death Master File (DMF) provided to the National Technical Information Service [Levin2019].

For our first launch of OMOP dataset, we limit the source data to Epic Clarity. The transformation process from Clarity to the OMOP-CDM can be summarized in five processes: ETL specifications, ETL code, mappings, data quality and data release. To design our ETL we current use OMOP-CDM v 5.3.1.The OHDSI consortium OMOP CDM github repository (https://github.com/OHDSI/CommonDataModel/releases/tag/v5.3.1) specifies a list of all the tables present in the 5.3.1 schema and the explanation for each one of the columns. The repository also contains a guide to populate each of the tables, THEMIS rules (https://github.com/OHDSI/Themis), which contain high level specifications for ETL. Those rules specify among other things a criteria for patient and clinical events inclusion and, what information is expected at each one of the fields. As OMOP is a patient-centric model, a patient is only brought in, iff, a patient has at least one qualified clinical event in Clarity. An event is considered a qualified event if it actually happened to the patient. A medication that was ordered but never administered to the patient will not be brought in into the CDM. There are also a large number of note records, ~51 million, that do not contain any text and are suspected to be EHR artifacts such as failed attempts to create a clinical note that get registered in the Epic database. These are excluded from OMOP ETL.

In Table S5.1, we present NCATS CTSA Common Metrics (https://ncats.nih.gov/ctsa/projects/common-metrics) from CTSA Program's coordinating center, Center for Leading Innovation and Collaboration (CLIC). These metrics were established as a common framework across the consortium of 58 CTSA centers in order to maximize the consortium's impact. Our OMOP database shows that over 60% of the patients have a diagnosis (ICD 9/10), over 40% have medication information (RxNorm), ~75% have lab information (LOINC), and over 90% of patients have clinical notes.

| | Data Domain | Count of patients (in millions) | % of patients |
|---|---|---|---|
| 1 | Total patients in the database = Unique patients with at least one encounter and at least one event | 2.7 | |
| 2 | Unique patients with at least one encounter and at least one event and administrative sex values | 2.7 | 100% |
| 3 | Unique patients with at least one encounter and at least one event and age or DoB | 2.7 | 100% |
| 4 | At least one encounter | 2.7 | 100% |
| 5 | At least one diagnosis (an ICD 9/10 or SNOMED value) | 1.8 | 67% |
| 6 | At least one procedure (an ICD 9/10 or CPT) | 2.6 | 95% |
| 7 | At least one medication (RxNorm) | 1.3 | 46% |
| 8 | At least one measurement / lab result (LOINC) | 2.1 | 78% |
| 9 | At least one device exposure | 0.3 | 10% |
| 10 | At least one observation | 1.9 | 69% |
| 11 | At least one notes / narratives | 2.5 | 91% |

Table S5.1: NCATS CTSA Common Metrics output

One of the main efforts to transform EHR data, Clarity, to the OMOP-CDM is the mapping of all the clinical information into standardized codes in the OHDSI OMOP vocabulary. The OHDSI OMOP vocabulary is a compendium of multiple terminologies and ontologies similar to the UMLS metathesaurus [UMLS2009]. The OMOP 5.3.1 vocabulary is organized across 39 domains, 371 classes and 430 relationships containing more than 6 million codes from 91 -vocabularies. By working with standardized codes, the ambiguity around specific use of strings, values or codes is reduced. The mapping process implies the translation of raw information (e.g. a string such as "ml") or codes in controlled terminologies or vocabularies (e.g. ICD10 R49.1) to standard codes in the OHDSI vocabulary. Figure S5.1 and table S5.2 show the improvement in standardized codes achieved over the raw Clarity data received from the Hospital. In some cases, Research IT had to develop a large number of custom codes (e.g. device exposure).

| | Unique Concept IDs | | |
|---|---|---|---|
| cdm_table | Standard (from Clarity) | Standard (by RIT) | Standard and Custom (by RIT) |
| Device_exposure (SNOMED) | 10 | 114 | 1,455 |
| Measurement (LOINC) | 4,783 | 3,040 | 3,040 |
| Condition_occurrence (an ICD 9/10 or SNOMED) | 14,814 | 3,359 | 3,359 |
| Procedure_occurrence (ICD 9/10 or CPT) | 28,204 | 2,382 | 2,382 |
| Drug_exposure (RxNorm) | 9,444 | 2,567 | 3,205 |

Table S5.2: Unique standardized codes delivered by Hospital Clarity and augmented by Research IT's ETL pipelines.

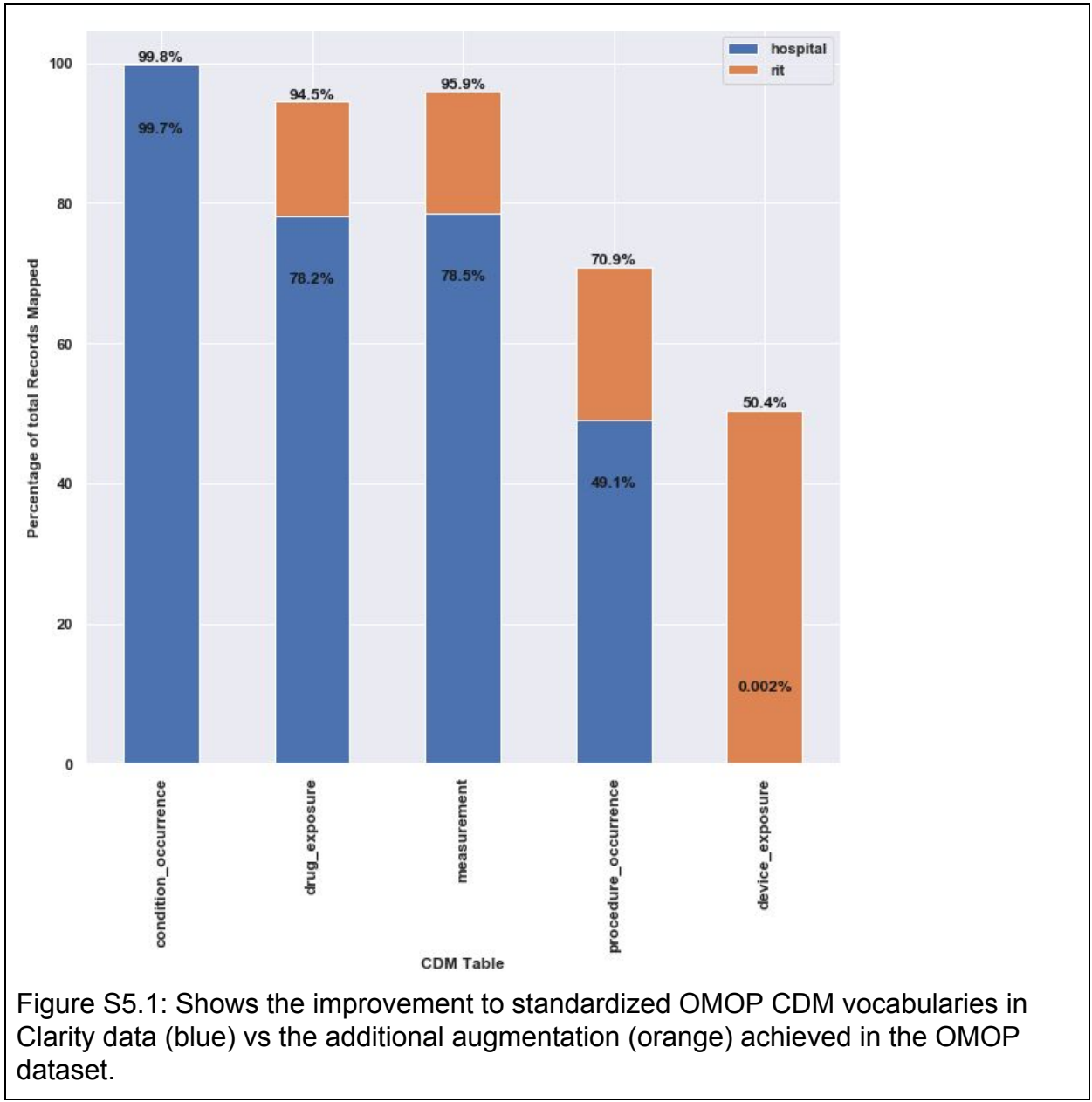

Figure S5.1: Shows the improvement to standardized OMOP CDM vocabularies in Clarity data (blue) vs the additional augmentation (orange) achieved in the OMOP dataset.

We use the OHDSI Achilles Heel tool for data quality analysis. This tool precomputes ~170 data characterizations that are iteratively used to improve the CDM. The improvements trigger three different actions: record removal, bug fixing and mapping changes.  In an example of record removal, we have cases where some clinical events occurring prior to birth date. This is essentially a data entry error where proper adjudication is not feasible, and as a consequence the only option is to remove the record from the CDM. Mapping changes occur in two scenarios. The first is when the source data does not contain the mappings and it is necessary to do custom mapping given the raw data (e.g. a string). The second is when a different table in the source Clarity data is needed - the two Hospitals have different workflows resulting in different tables being populated in Clarity. This data quality process integrates with all the transformation by possibly modifying the specs and the subsequent code. It is essentially an iterative process of continuous improvement.

Figure S5.2 shows the distribution of words in notes. There are `~100M notes containing 33 billion words. Median number of words in a note is ~100, and ~9% notes have >1000 words, and 0.04% of the notes have >5,000 words. We investigated the notes that contained >5,000  words and found a few different reasons for their uncommon length. In some cases,  the long notes include information of laboratory tests, radiology reports, medications and checklists beyond the regular use of those elements in a note. Discharge summaries sometimes contain information that is already present in progress notes as a way to document the progression of the patient during the stay. In some imaging reports, we found detailed information about the anatomy observed, specifically for cases where multiple systems or organs are being evaluated. Lastly, one of the EHR migrations resulted in merging of notes for a subset of our patients.

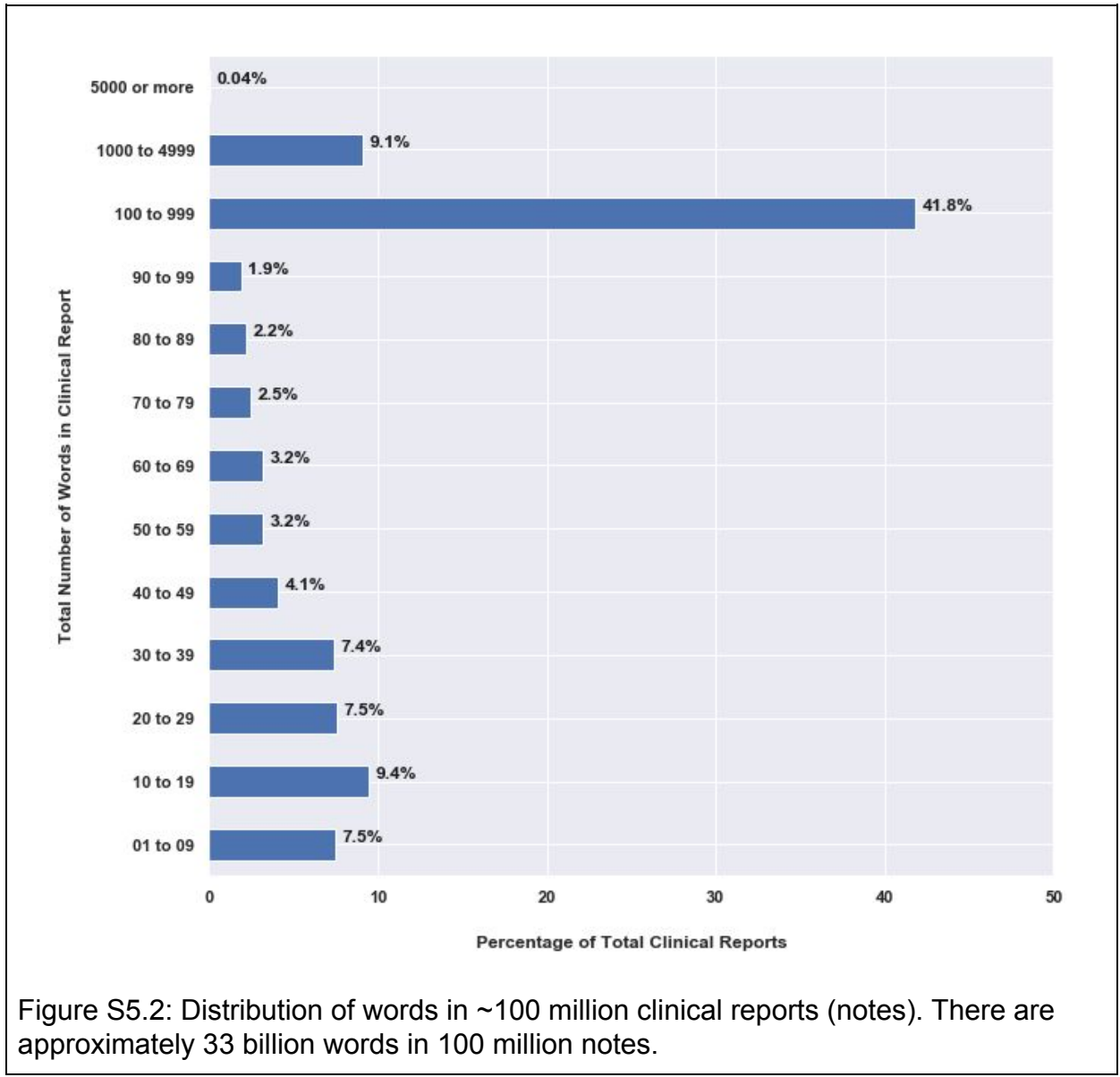

Figure S5.2: Distribution of words in ~100 million clinical reports (notes). There are approximately 33 billion words in 100 million notes.

## Section 6: De-identification methods

In this section, we present anonymization techniques used to produce the STARR-OMOP-deid dataset. The anonymized clinical text is stored in NOTE table of the OMOP CDM.

Research IT achieves de-identification of STARR-OMOP data in accordance with NIST guidelines [NISTIR 8053] to meet the HIPAA Privacy Rule [HHS2002]. In particular, we use the Safe Harbor approach. Our approach to de-identification is similar to those presented by Ferrández *et al* [Ferrández2013] in that, we a) maximize patient confidentiality by redacting as much PHI as possible and may accidentally redact non-PHI; and b) leave de-identified data in a usable state preserving as much clinical information as possible.

Prior assessments [Fernandez2013] of clinical text de-identification techniques show that it is difficult to find a single approach that performs well in all cases. Clinical narratives can be fragmented and lack formatting, making the use of pre-trained traditional newswire Name Entity Recognition (NER) approaches limiting. Rule based techniques (e.g. pattern matching) are not scalable approaches. Research IT has developed a de-identification approach, TiDE (Text DEidentification), that combines a mix of pattern matching techniques, machine learning-based NER and Hiding in Plain Sight [Carrell2013]. While each of the techniques is only effective part of the time, together they are highly effective most of the time. We focus TiDE pipeline (Figure S6.1) on preserving patient privacy i.e., sensitivity of finding PHI is more important than specificity.  In the first step, we find the locations of HIPAA identifiers.

TiDE has three separate sub-modules that find the HIPAA identifiers, a) the NLP name entity recognition module uses CoreNLP [Manning2014]] and can find random combinations of street name, city name, state name and zip code, b) the regular expression (regex) pattern matching to known patient PHI and, c) for entities like MRNs, SSN (e.g. 123-45-6789), email (e.g. john@example.org), IP, URL, we use an enumerated pattern matching.

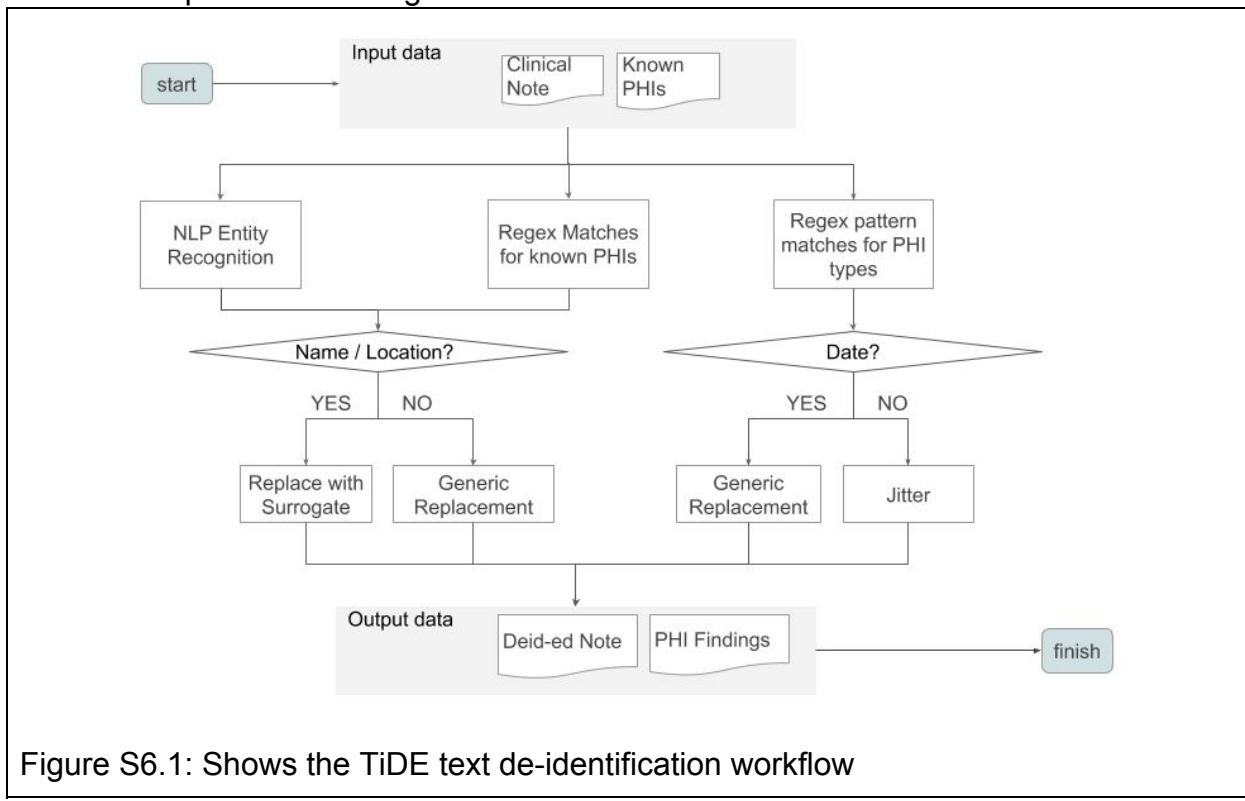

Figure S6.1: Shows the TiDE text de-identification workflow

We selected CoreNLP [Manning2014], an open source software developed at Stanford for identifying names and addresses in the text. At the heart of CoreNLP is the Stanford Named Entity Recognizer (Stanford NER) also known as CRFClassifier. The software provides a general implementation of linear chain Conditional Random Field (CRF) sequence models. CRFs have been known to perform well for clinical text de-identification [Deleger2013].

All HIPAA findings are stored in BigQuery. We then use BigQuery User-Defined Function (UDF, https://cloud.google.com/bigquery/docs/reference/standard-sql/user-defined-functions) to select and merge all findings across the different methods to generate final list of candidate PHI that are then de-identified.

| Original PHI & PII | Safe Harbor Lookup, NLP, Reg-Ex, Leaked PHI | TiDE Safe harbor, Surrogate PHI / PII, hiding PHI |
|---|---|---|
| Jonathan Smith, 41 years … children, Lynn and David and Madison … oncologist, Dr. White on 5/13/10 to schedule MRI … DRE from 5/7 was … to call Mr. Smith on ... | [**PAT-FN] [**PAT-LN], 41 years … children, [**NAME] and [**NAME] and Madison … oncologist, Dr. [**DR-LN] on [**5/31/10] to schedule MRI … DRE from 5/7 was … to call Mr. [**PAT-LN] on … | Tom Jones, 41 years … children, Mary and Joe and Madison … oncologist, Dr. Howe on 5/31/10 to schedule MRI … DRE from 5/7 was ... to call Mr. Jones on … |

Table S6: Illustration of our TiDE output.

For names and places, which are two of the three most frequent PHI in clinical text data, we use a surrogate approach [Carrell2013] aka Hiding in Plain Sight (HIPS) illustrated in Table S6. In column 1, we show original PHI (patient and children names, date of visits) and PII (doctor's name). In column 2, we show a typical Safe Harbor implementation, a) patient and doctor names are identified and redacted using pre-existing information in Clarity database, b) one of the dates is identified using RegEx and jittered, and b) two of the children names are identified using NLP approaches and redacted. In column 3, we show the impact of surrogate approach to names. Here, one of the names (Madison) and a procedure date (5/7), previously undetected, are hidden with the HIPS approach. At the time of writing, the name replacement is gender aware but not ethnicity aware. Addresses are replaced randomly.

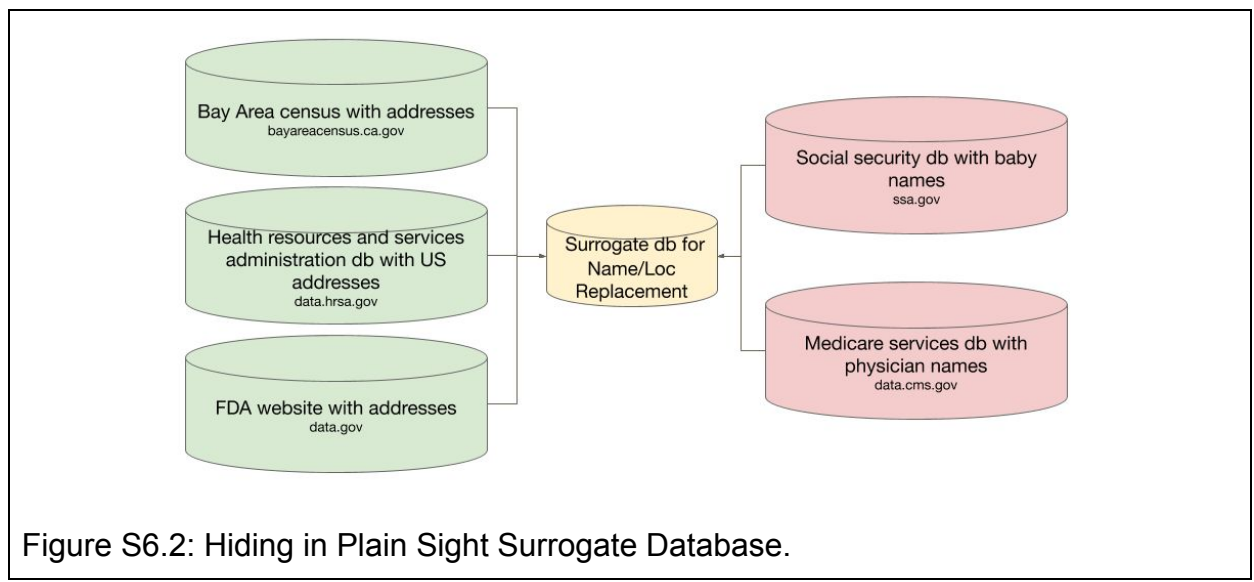

Figure S6.2: Hiding in Plain Sight Surrogate Database.

The development of surrogate database is illustrated in Figure S6.2. We collate the following sources, a) Bay Area census with addresses (bayareacensus.ca.gov), b) Health resources and services administration database with US addresses (data.hrsa.gov), c) FDA website with addresses (data.gov), d) Social security database with baby names (ssa.gov), and e) Medicare services database with physician names (data.cms.gov).

Recently, the HIPS team presented the parrot attack [Carrell2019] and for this attack to succeed, we will need to assume an attack scenario where a) the attacker will maliciously hack in our secure data center and download de-identified data, b) the attacker will attempt to tag all real and surrogate PHI in the data, c) attacker will have access to data used to train our models, and d) attacker will be able to train a model to match ours. We believe that our system does not support this attack scenario. Firstly, Research IT does not provide any training data and secondly, our pipeline uses a collection of different machine learning algorithms and Safe Harbor method *i.e.*, it is not a single model that can be reverse engineered.

We use a manual QC process illustrated in Supplementary Figure S6.3. In the case of clinical notes, we start with top 200 different note types that occur most frequently, pick 1000 random notes from these and of these, we pick a 100 that have a high amount of PHI found or high number of words found. These hundred go through a manual review for false negatives *i.e.*, we look for PHI left behind. For flowsheet (observational measurements) data, we take 80,000 rows and take them through a word frequency count. We then manually review the lowest frequency 10,000 words for potential PHI.

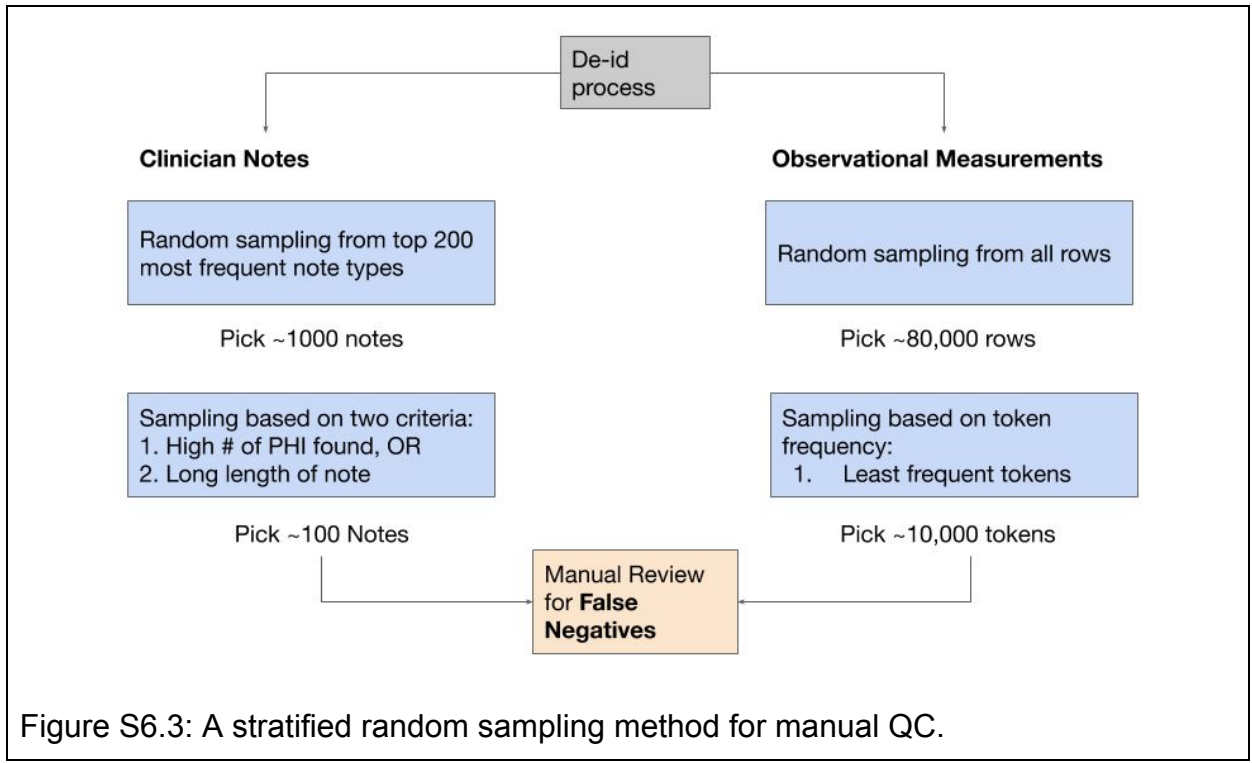

Figure S6.3: A stratified random sampling method for manual QC.

We considered using the i2b2 2014 *Deidentification & Heart Disease* dataset [Stubbs2014] to evaluate the performance of our de-identification pipeline but we could not find a meaningful set of comparison metrics due to procedural differences between the two. Firstly, we heavily use *a priori* information for a given patient. This information is not available for the i2b2 dataset. Secondly, our interpretation of the HIPAA Safe Harbor method is similar to the one used in MIMIC III [Johnson2016] and is different from the i2b2 dataset. For example, in the i2b2 dataset, *age* is considered PHI. However, in our case and in MIMIC III, only ages >89 are considered PHI. Other categories with difference from i2b2 approach include *profession* and *location*. Profession only becomes an identifier when it is very unique (e.g. Apple CEO). Location for i2b2 dataset includes organization as PHI and not just addresses. As with profession an organization becomes an identifier only if it is very unique and can be easily linked with the patient. In our strategy we used CoreNLP to recognize generic locations and organizations, however, by no means is CoreNLP capable of removing mentions of obscure company names.

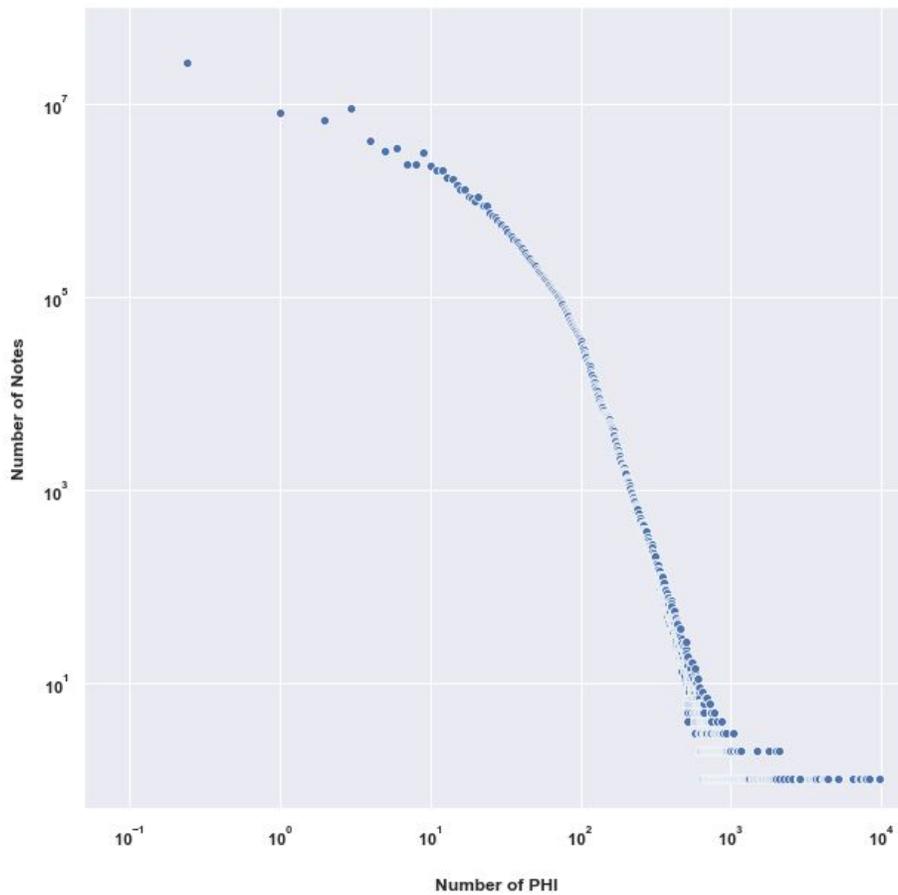

Figure S6.4: Shows distribution of PHI count in ~100 million notes.

Figure S6.4 shows the distribution of PHI count found in notes. There are ~22 million notes with no PHI findings, and ~1.3M notes with more than 100 PHI findings. There are ~33 billion words, of which 1.4 billion are determined as PHI findings *i.e.*, approximately 4% of the words are found to be PHI. Some of the notes with high volumes of PHI are due to the presence of repeated elements inside checklists produced by dropdown menus, laboratory results and vital signs. The most common repeated elements are dates and clinician's names.

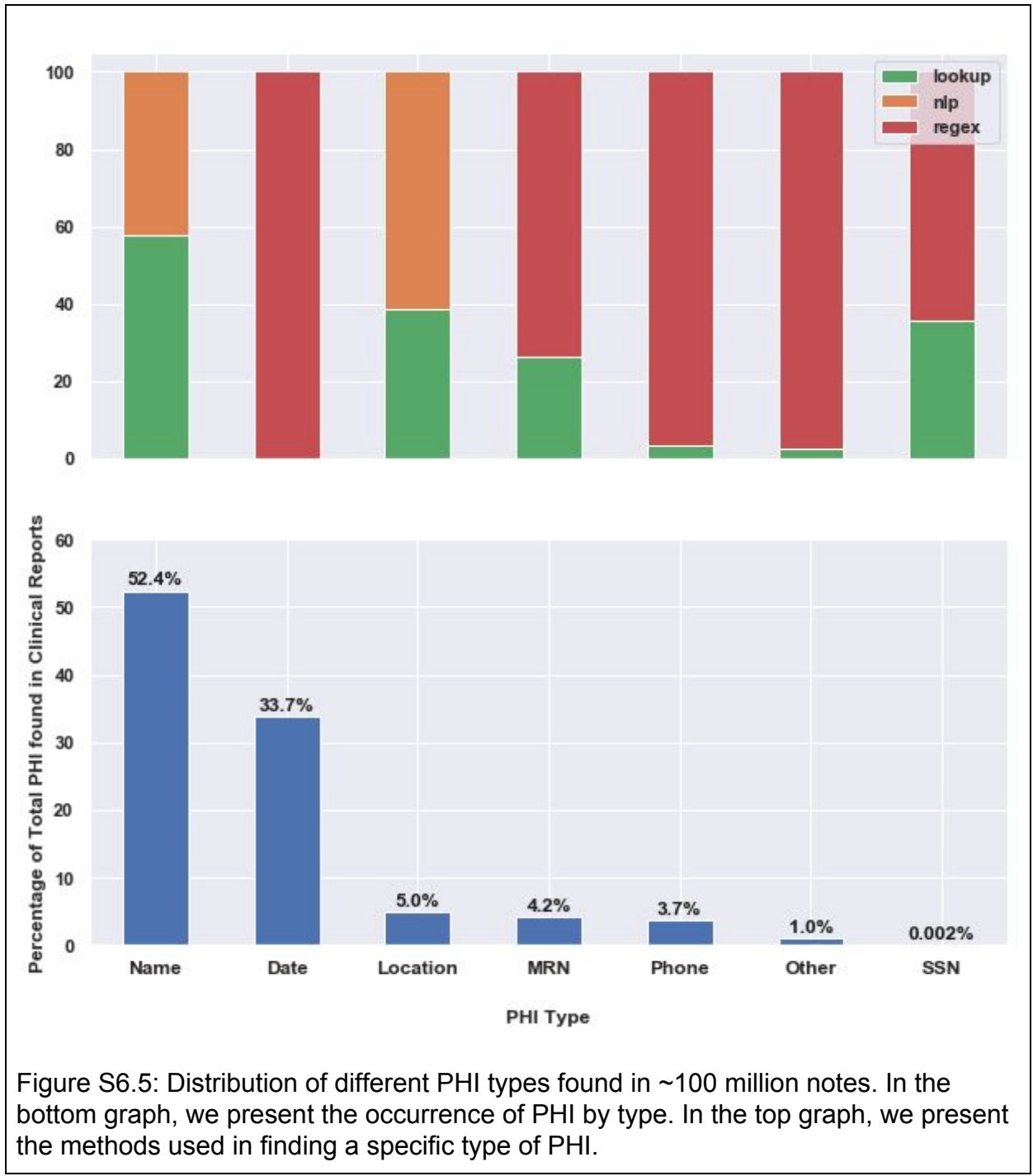

Figure S6.5: Distribution of different PHI types found in ~100 million notes. In the bottom graph, we present the occurrence of PHI by type. In the top graph, we present the methods used in finding a specific type of PHI.

Figure S6.5 presents findings from our text anonymization pipeline. We observe that names, dates, and location are the most common PHI types. We also find that the "mixed bag" approach of finding PHI is useful. For example, for patient and physician names, ~60% are found using patient information lookup and an additional 40% are found using NLP. Dates, MRNs and phone numbers benefit from regex search.

# Section 7: Processing clinical text to identify known medical concepts

In this section, we present our method used to process clinical text to identify known medical concepts and store these concepts in the NOTE_NLP table in the OMOP CDM.

Text processing tools transform the unstructured information inside the notes into structure information that can be queried and stored in structured tables. The amount and type of information extracted from clinical notes vary from tool to tool. Most of the tools focus their efforts in mapping mentions of diseases, signs, and symptoms into a controlled terminology such as ICD or SNOMED  (Figure S7.1). Through this mapping two different mentions such as "fever" and "temp > 37.5C" are mapped to the same concept inside a terminology. In this case the ICD-10 code R50.9.

For populating concepts in the NOTE_NLP table, we use a pipeline developed by LePendu *et al* [LePendu2013], that has incorporated both negation detection and history detection. These contextual cues are based on NegEx [Chapman2001] and ConText [Chapman2007] and enable us to discern whether a term should *not* be attributed the current patient (e.g., *lack of* valvular dysfunction, or *sister has* muscular dystrophy).  The pipeline extracts term mentions *i.e.*, string unique identifiers (SUI) within the UMLS [RRF2009] vocabulary. The terms are disambiguated (Figure S7.2) and then mapped to concepts in the OHDSI vocabulary (Figure S7.3) using an ETL similar to the OHDSI tool Ananke (https://github.com/jmbanda/Ananke) which maps from CUI to concepts. There are two ways to disambiguate via pre-processing, a) Word sense disambiguation algorithms that attempt to select the most likely sense of a string given the context, and b) the dictionaries used in the mapping are pruned and filtered using rules that attempt to increase precision.  STARR uses the latter approach, we exclude terms with less than four characters and those present in the list of ambiguous terms provided by UMLS  in AMBIGSUI.RRF.

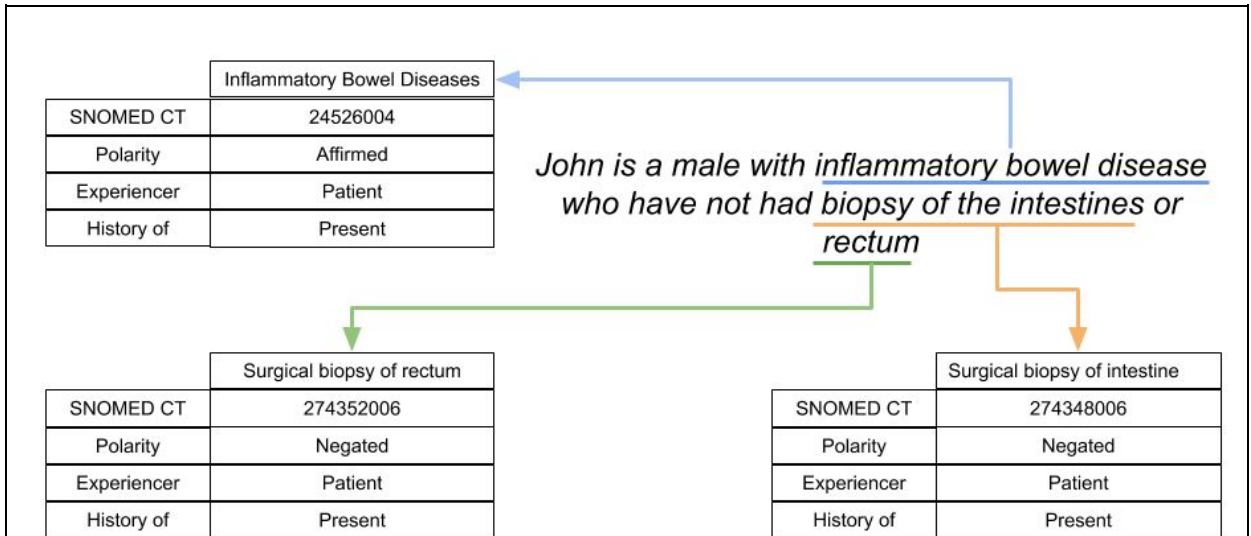

Figure S7.1: An example of the mapping of term mentions of diseases and procedures from clinical notes into a controlled terminology such as SNOMED.

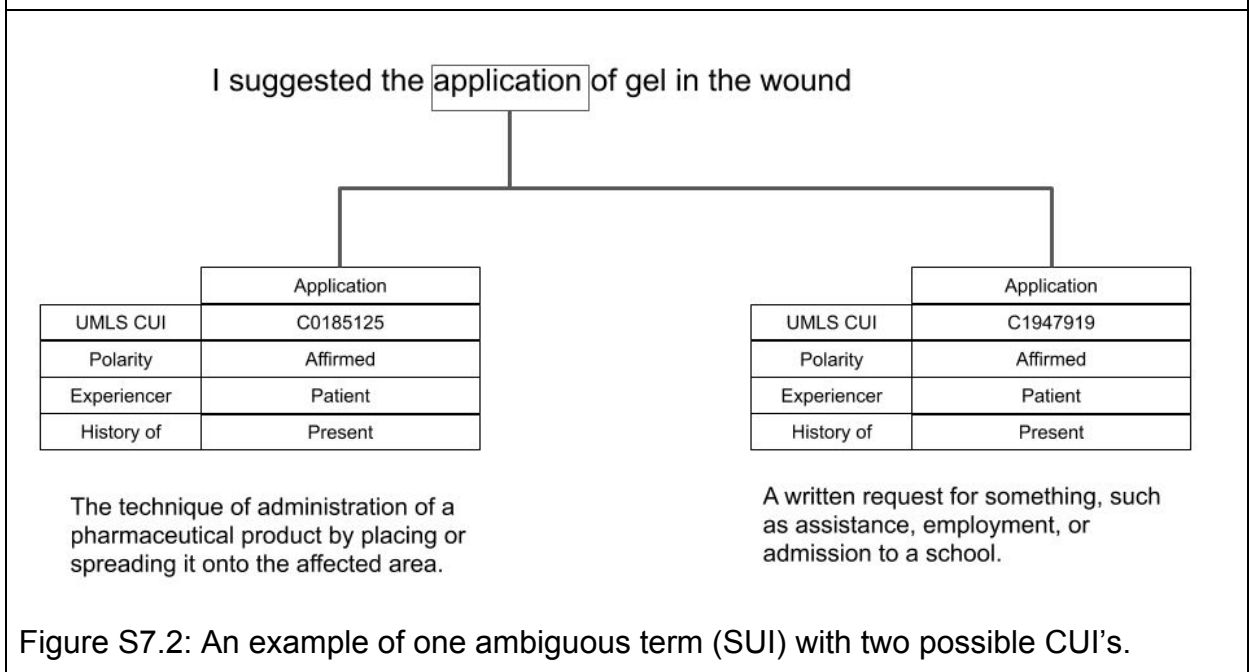

Figure S7.2: An example of one ambiguous term (SUI) with two possible CUI's.

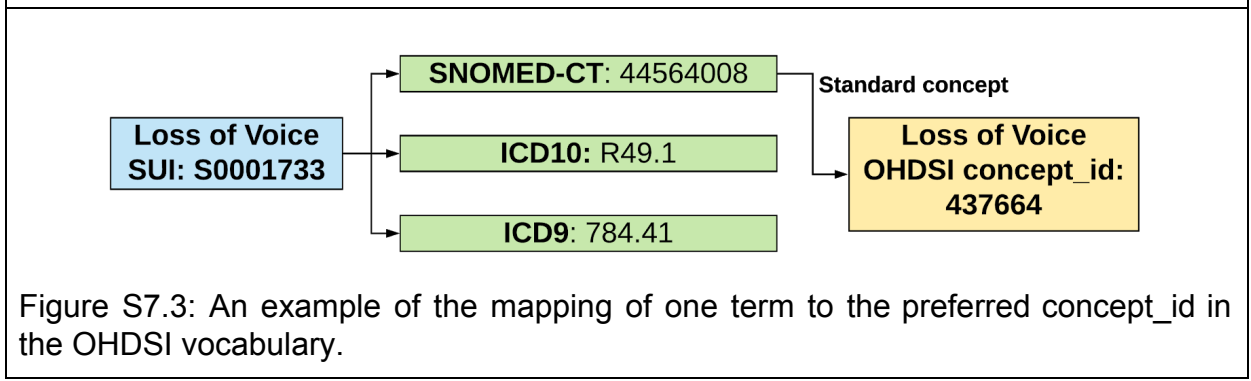

Figure S7.3: An example of the mapping of one term to the preferred concept_id in the OHDSI vocabulary.

In the pipeline developed by LePendu *et al* [LePendu2013], each concept can contain modifiers that affect their interpretation and use in research. A mention of "high blood pressure" does not ascertain that the patient is experiencing the symptom. Context is necessary for an adequate interpretation [Harkema2009]. If the sentence, "*family history* of high blood pressure", human intelligence can safely interpret that the patient is not the one that suffers from high blood pressure. To translate this into structured information we need to introduce concept modifiers. These modifiers allow us to enhance the interpretation of the mapped concepts.

We populate the term_modifiers field in the NOTE_NLP table using the following strings:

1. experiencer_other: The absence of this modifier implies that the clinical event associated with the concept is being experienced by the patient. In the two examples, a) "The *patient* has coronary artery disease" and, b) "Coronary artery disease in *father*", the modifier is present only in example b.

2. history_of_past: If this modifier is present, the clinical event associated with the concept likely occurred in the past. In the three examples, a) "she *has* chest pain", b) "She *had* hyperlipidemia" and, c) "*Will* schedule screening mammogram at *next visit*"; the modifier is present only in example a.

3. polarity_negated: The absence of this modifier implies the concept is present for the patient. If this modifier is present, the concept is negated. In the two examples, a) "the patient *does not* have any neurologic deficits" and, b) "*she is being* dialyzed", the modifier is present only in example a.

Table S7 shows the distribution of concepts from different vocabularies and domain found in clinical notes respectively. Column 2 shows total counts of the concepts found (sum = 1.9 billion) and column 3 shows the percentage of occurance. Column 4 shows total unique counts (sum = 122K) and column four shows the percentage of unique count occurance.

| | Vocabulary | Total count (in millions) | % of Total count | Total unique concept count | % of total unique concept count |
|---|---|---|---|---|---|
| 1 | SNOMED | 1,179 | 60.99 | 68,932 | 56.55 |
| 2 | RxNorm | 47 | 2.46 | 12,755 | 10.47 |
| 3 | LOINC | 407 | 21.03 | 8,008 | 6.57 |
| 4 | ICD10CM | 9.6 | 0.50 | 7,113 | 5.84 |
| 5 | MeSH | 116 | 5.79 | 6,745 | 5.53 |
| 6 | NDFRT | 87 | 4.50 | 4,796 | 3.94 |

| 7 | ICD9CM | 8 | 0.41 | 3,688 | 3.03 |
| 8 | CPT4 | 9.5 | 0.49 | 3,302 | 2.71 |
| 9 | Multum | 34 | 1.77 | 2,971 | 2.44 |
| 10 | SNOMED Veterinary | 13 | 0.68 | 1,309 | 1.07 |
| 11 | HCPCS | 1.7 | 0.09 | 878 | 0.72 |
| 12 | ICD10 | 0.9 | 0.05 | 690 | 0.57 |
| 13 | ATC | 24 | 1.22 | 523 | 0.43 |
| 14 | ICD10PCS | 0.1 | 0.01 | 125 | 0.10 |

Table S7: Frequency of occurrence of top 14 vocabulary terms found as a result of text mapping.

In the case of vocabularies, preference to SNOMED in OMOP is reflected in the data. The prevalence of vital signs and laboratory results in clinical notes, is reflected in LOINC being the second most common vocabulary. MeSH vocabulary includes the subject descriptors appearing in MEDLINE/PubMed, the NLM catalog database, and other NLM databases. The vocabularies NDFRT, RxNorm, Multum and ATC cover the mentions of drugs and chemicals within the clinical notes.  Two other vocabularies, CVX and NUCC, are found to have insignificant counts.

## Section 8: Nero, a secure data science platform

In this section, we present our secure data science platform from where STARR-OMOP-deid can be accessed and analyzed.

The approach of co-locating data with computing resources and analysis tools is popularized in platforms such as NCI Genomics Data Commons [Wilson2017] and Sage Synapse Platform [Ellrott2019]. Academic centers have also taken the initiative of developing Big Data science platforms [McPadden2019]. There are multiple secure and compliant workspaces offered by commercial entities (e.g. DNAnexus, https://www.dnanexus.com/) and Supercomputing facilities (e.g. San Diego Supercomputer Center Sherlock Cloud, https://sherlock.sdsc.edu/), but use of these facilities for High Risk or PHI data requires lengthy contractual processes. We believe that a secure internal infrastructure, that is integrated with institutional compliance processes, leads to reduced friction for researchers.

To support the High Risk nature of the underlying data, Stanford has built a PHI secure data science platform, Nero (http://med.stanford.edu/nero), named after the fictional armchair detective Nero Wolfe (https://en.wikipedia.org/wiki/Nero_Wolfe). The access to STARR-OMOP-deid dataset is made available in Nero. The platform is a collaboration between Research IT at Stanford Medicine and Stanford Research Computing Center (SRCC) at University IT. The SRCC team has in-depth expertise across the spectrum of HPC and data management services. The SRCC team develops, and manages the Nero platform and provides secure workspaces and research computing support to the users.  Nero has a private cloud infrastructure based on open source components hosted at Stanford Research Computing Facility (SRCF) with direct connectivity to a 100 gigabit network connection to available Research and Education Networks via CENIC's California Research and Education Network (https://cenic.org/). Nero is also integrated with Stanford secured public clouds. The platform allows researchers to access Stanford data resources like STARR and also allows researchers to bring their own data to the platform.

The on-premise Nero environment, at the time of writing, is comprised of 6 Dell C4140 and 18 Lenovo SD530 servers. Servers have either dual Intel Xeon Silver or dual Intel Xeon Gold processors, providing 24 cores with 384 GB of RAM per server.  These compute nodes use Ubuntu running Docker containers, with relevant software. Workloads are scheduled using Slurm Workload Manager (https://slurm.schedmd.com/documentation.html) , and configurations are managed by Metal as a Service (MAAS, https://maas.io/), Juju as a Service (JaaS, https://jaas.ai/), and Kubernetes container orchestration service (https://kubernetes.io/).  Nero also has 24 NVidia V100 Tensor Core (https://www.nvidia.com/en-us/data-center/v100/) and 4 NVidia Tesla P100 (https://www.nvidia.com/en-us/data-center/tesla-p100/) GPUs totaling 28 GPU cores available to researchers.  For storage, Nero has 630 TB of Ceph (https://docs.ceph.com/docs/master/) general-purpose storage on Lenovo storage arrays, plus 30 TB of high-IOPS storage.  Storage is encrypted via Linux Unified Key Setup-on-disk-format (or LUKS) Advanced Encryption Standard (AES) 265-bit encryption.

Finally, Nero makes commonly used research tools available using two approaches, a) for publicly available tools, the Nero team containerize the application stack and take it through a rigorous process of vulnerability scanning (e.g. OHDSI analytical tools), and b) for software-as-a-service (SaaS), the team partners with the provider to integrate the service with the Nero platform (e.g. https://stanford.md.ai/) in compliance with Stanford ISO requirements. The system administration of Nero platform includes a fully managed security service and research computing support.

## Section 9: Data and tool access

In this section, we present how researchers access data and data science tools in the secure Nero environment.

A researcher can request access to Nero as long as they are affiliated with a Stanford Principal Investigator (PI) and have a Stanford identity. Once the researcher becomes an authorized user, they get access to Nero via Stanford Virtual Private Network (VPN) and Stanford two factor authentication. Researcher needs to access Nero using a secure Stanford endpoint (*i.e.*, laptop) that meets minimum security requirements for High Risk and PHI data. Stanford researchers are also required to take HIPAA training (https://privacy.stanford.edu/training/hipaa). To access STARR-OMOP-deid data,

Stanford researcher then completes the requisite Data Use Agreement, also referred to as Data Privacy Attestation, attesting to the fact that they are requesting access to a High Risk non-human subject exploratory dataset. The attestation goes through the researcher's responsibilities for the data handling. Once a researcher is on Nero, and DUA is completed, they get access to the STARR-OMOP-deid BigQuery dataset and are added to a user support slack channel.

Once on Nero, they can use an interactive terminal or use Jupyter Hub to launch a preferred kernel (e.g. R or Python) or use batch mode via SLURM job scheduler. They can access their workspaces on-premise or Cloud and their preferred applications, bring their own data to their workspace, run batch mode or real time analysis, request new applications or troubleshooting help from the system administration team, and share code with collaborators using Stanford gitlab.

In Table S9.1, we present performance results from common queries as a result of using Google BigQuery to store the OMOP data (Supplementary material, Section 3). OHDSI development community has made an explicit effort to support a number of databases including traditional relational data warehousing technologies such as PostgreSQL, Oracle, Microsoft SQL, as well as Big Data warehousing technologies such as Azure Synapse, IBM Netezza, and Amazon RedShift, Cloudera Impala and Google BigQuery. The OHDSI SqlRender package (https://github.com/OHDSI/SqlRender) and JDBC drivers require performance tuning to support a new database type. RIT worked closely with Odysseus Data Services Inc, OHDSI consortium's ATLAS development team, to optimize performance of Google Big Query for OHDSI ATLAS tool (Supplementary Section 3). These queries are from the OHDSI Achilles tool (https://github.com/OHDSI/Achilles/tree/master/inst/sql/sql_server) and are used to characterize the data quality. Out of 725 total queries available in Achilles, 660 queries took less than 17 seconds, and median execution time was 3 sec. The longest query took 1207 seconds (20 min) and processed 13GB of data.

| | |
|---|---|
| Number of Queries | 1.0 |
| Mean Execution Time (s) | 9.4 |
| Median Execution Time (s) | 2.8 |
| Min Execution Time (s) | 0.4 |
| Max Execution Time (s) | 1,207.4 |
| Total Execution Time (hr) | 1.9 |
| Volume of data traversed by the queries (GB) | 726.3 |
| Source of Queries | Achilles Analyses Queries |

Table S9.1: We run OHDSI Achilles QA SQL queries (725) and report the statistical query times.

Researchers also get access to OHDSI ATLAS Cohort tool (https://ohdsi-atlas.stanford.edu/). OHDSI support for BigQuery platform is relatively new. Research IT has worked closely with Odysseus Data Services Inc, developers of OHDSI ATLAS cohort analysis tool, to optimize ATLAS performance for BigQuery. We present the results of performance benchmarking using postgreSQL and BigQuery for the commonly available Medicare Claims Synthetic Public Use datafiles (DE-SynPUF, https://www.cms.gov/Research-Statistics-Data-and-Systems/Downloadable-Public-Use-Files/SynPUFs/DE_Syn_PUF). In order to benchmark postgreSQL vs BigQuery, ATLAS was installed on Google Compute Engine instance with parameters presented in Table S9.2. We used OHDSI CDM DDL scripts to build the two data structures - PostgreSQL (https://github.com/OHDSI/CommonDataModel/tree/v5.3.1/PostgreSQL) and BigQuery (https://github.com/OHDSI/CommonDataModel/tree/v5.3.1/BigQuery). In case of PostgreSQL indexes and constraints were applied. For BigQuery, no additional configuration was applied. Standard suite of tests used for ATLAS benchmarking (https://github.com/odysseusinc/TestScripts/tree/master/scripts), were run. Each test was executed separately per database to avoid caching and resource contention. Results of the various tests are presented in Table S9.3.

| ATLAS compute instance configuration | Deployed ATLAS 2.7.4 with 2 vCPUs, 15 GB memory, 50Gb storage, and CentOS 7 |
|---|---|
| DE-SynPUF dataset | OMOP CDM ver 5.3.1, was used with 2M patients for measurement with vocabularies from ATHENA October 2019 |
| PostgreSQL VM instance | Deployed PostgreSQL v11, latest available in GCP with 4 vCPUs, 15 GB memory, 903 GB SSD storage |
| PostgreSQL configuration | temp_file_limit: 2,147,483,647, max_connections: 40, maintenance_work_mem: 960,000, checkpoint_completion_target: 0.7, default_statistics_target: 100, random_page_cost: 1.1, work_mem: 48,000, max_wal_size: 2,100,000,000. |

Table S9.2: ATLAS benchmarking setup on GCP.

| # | Action | PostgreSQL (SynPUF 2m) | BigQuery (SynPUF 2m) |
|---|---|---|---|
| 1 | Vocabulary search (aspirin) | 00:00:17 | 00:00:17 |
| 2 | Profile search (2320055) | 0:00:03 | 00:00:25 |

| 3 | Cohort generation (counts): Celecoxib new users (Target Cohort) | 00:08:08 | 00:02:21 |
| 4 | Characterization generation: Aspirin users vs Clopidogrel users characterization | 00:10:20 | 00:07:38 |
| 5 | IR generation: Incidence rate valsartan tutorial | 00:16:39 | 00:04:35 |
| 6 | PLE: Celecoxib vs Diclofenac in major GI bleeding | 12+ hours | 222m 43s |
| 7 | PLP: Clopidogrel leading to pneumonia | 12+ hours | 87m 09s |
| 8 | Cohort Pathway: PNAS Characterizing treatment pathways replication | 00:38:35 | 00:10:14 |

Table S9.3: ATLAS benchmarking results using a synthetic dataset, SynPUF.

## Section 10: User training

In this section, we present the training material available to our community.

As part of data access, researchers also get access to technical specification documents, training datasets - a one percent STARR-OMOP-deid dataset (for testing SQL queries ) and a synthetic dataset Data Entrepreneurs' Synthetic Public Use File (DE-SynPUF, https://www.cms.gov/Research-Statistics-Data-and-Systems/Downloadable-Public-Use-Files/SynPUFs/DE_Syn_PUF.html) in OMOP 5.3.1 format, and access to Stanford gitlab with Jupyter notebooks compatible code. The specifics of ETL code used to transform data from Clarity to OMOP is made available upon request.

Research IT offers hands-on training sessions that help researchers with understanding and use of OMOP CDM, STARR-OMOP-deid dataset, and the Nero research computing environment.